\RequirePackage{silence}
\documentclass[Acad.Quant.,researcharticle,9pt,twoside,twocolumn,oneauthor]{jams_academia}

\newcommand{\ecedilla}{e\kern-1.2pt\rotatebox{180}{\raisebox{-0.205ex}{\scriptsize c}\kern-1.85pt}}

\usepackage{subcaption}
\usepackage{tikz}
\usetikzlibrary{shapes.misc,arrows,positioning}
\usepackage{quantikz}

\usepackage{tabularx,array}

\usepackage{etoolbox}
\apptocmd{\sloppy}{\hbadness 10000\relax}{}{}

\AtBeginDocument{\RenewCommandCopy\qty\SI}
\ExplSyntaxOn
\msg_redirect_name:nnn { siunitx } { physics-pkg } { none }
\ExplSyntaxOff

\makeatletter
\renewcommand{\maketag@@@}[1]{\hbox{\m@th\normalsize\normalfont#1}}
\makeatother

\newcommand*{\N}{\mathbb{N}} 
\newcommand*{\z}{\mathbb{Z}} 
\newcommand*{\Q}{\mathbb{Q}} 
\newcommand*{\R}{\mathbb{R}} 
\newcommand*{\C}{\mathbb{C}} 
\newcommand*{\hil}{\mathcal{H}} 
\newcommand*{\one}{\mathds{1}} 

\newcommand*{\lo}{\mathcal{L}} 

\newcommand*{\defcolon}{\,:\,}
\newcommand*{\bits}{Z(N)}
\newcommand*{\COP}{\mathcal{C}} 

\DeclareMathOperator{\spn}{span}
\DeclareMathOperator{\OSSP}{OSSP}
\DeclareMathOperator{\Aut}{Aut}
\DeclareMathOperator{\Times}{times}
\DeclareMathOperator{\Sym}{Sym}
\DeclareMathOperator{\Phase}{P}
\DeclareMathOperator{\Mix}{M}
\DeclareMathOperator{\SWAP}{SWAP}
\DeclareMathOperator{\id}{id}

\newcommand*{\phaseu}{U_{\Phase}}
\newcommand*{\mixu}{U_{\Mix}}
\newcommand*{\simmixu}{U_{\Mix, 0}}
\newcommand*{\seqmixu}{U_{\Mix, \sigma}}

\newcommand\scriptin{\raisebox{0.15ex}{$\scriptscriptstyle\in$}} 

\firstpage{1}
\pubvolume{2}
\issuenum{3}
\pubyear{2025}
\copyrightyear{2025}
\externaleditor{Academic Editors: Misha Erementchouk, Yuhua Duan}
\datereceived{2025-04-11}
\dateaccepted{2025-08-29}
\datepublished{2025-09-12}
\doinum{10.20935/AcadQuant7900}
\Title{Symmetry-based quantum algorithms for open-shop scheduling with hard constraints}

\TitleCitation{Symmetry-based quantum algorithms for open-shop scheduling with hard constraints}

\Author{Lennart Binkowski$^{1}$, Gereon Ko{\ss}mann$^{2,3}$, Christian Tutschku$^{3}$, Ren\'{e} Schwonnek$^{1,}$*}

\AuthorNames{Binkowski L, Ko{\ss}mann G, Tutschku C, Schwonnek R}

\AuthorCitation{Binkowski L, Ko{\ss}mann G, Tutschku C, Schwonnek R}

\address{%
    \textsuperscript{1}{Institute for Theoretical Physics, AG Quantum Information, Leibniz University Hannover, Hannover, Germany.}\\
    \textsuperscript{2}{Institute for Quantum Information, RWTH Aachen University, Aachen, Germany.}\\
    \textsuperscript{3}{Fraunhofer Institute for Industrial Engineering IAO, Stuttgart, Germany.}
}

\corres{rene.schwonnek@itp.uni-hannover.de}

\abstract{Encoding hard-constrained optimization problems into a variational quantum algorithm often turns out to be a challenging task.
    In this work, we provide a solution for the class of open-shop scheduling problems (OSSPs), which we achieve by rigorously employing the symmetries of the classical problem.
    An established approach for encoding the hard constraints of the closely related traveling salesperson problem (TSP) into mixer Hamiltonians was recently given by Hadfield et al.'s Quantum Alternating Operator Ansatz (QAOA).
    For the OSSP, which contains TSP as a special case, we show that desired properties of similarly constructed mixers can be directly linked to a purely classical object: the group of feasibility-preserving bit value permutations.
    We also outline a generic way to construct QAOA-like mixers for these problems.
    We further propose a new variational quantum algorithm that incorporates the underlying group structure more naturally and, as a proof of principle, implement our new algorithm for a small OSSP instance on an IBM Q System One.
    Unlike the generic QAOA, our algorithm allows for bounding the amount and the domain of parameters necessary to reach every feasible solution from above: Optimizing at most quadratically many parameters should suffice to reach the optimum with certainty.}


\begin{document}

\section{\label{section:Introduction}Introduction}

Logistic and scheduling tasks are a major branch within the collection of hard optimization problems for relevant industrial applications.
A prominent representative of those is the \textit{open-shop scheduling problem} $\text{OSSP}(M, T, J)$, which we consider in this work:
Given $M$ machines with $T$ time slots each, one has to distribute $J$ jobs such that every job gets performed precisely once and no position is filled with more than one job.
Not only is the OSSP at the mathematical core of many real-world problems, it also prominently incorporates the well-known \textit{traveling salesperson problem} (TSP) as a subclass.
More specifically, it holds that $\text{TSP}=\text{OSSP}(1,T,T)$.

{\addfontfeature{LetterSpace=0.5}Its practical relevance as well as the fact that solving an instance of OSSP can easily turn out to be a hard task for classical computers, make it an interesting target for the application of quantum algorithms.
    Confronted with the restricted capabilities of available quantum computer architectures, the class of \textit{variational quantum algorithms}~\cite{Cerezo2021VariationalQuantumAlgorithms,Marshall2020CharacterizingLocalNoiseInQAOACircuits,Peruzzo2014AVariationalEigenvalueSolverOnAPhotonicQuantumProcessor,Barboutsos2020ImprovingVariationalQuantumOptimizationUsingCVaR,Larkin2022EvaluationOfQAOABasedOnTheApproximationRatioOfIndividualSamples} (VQAs) is receiving a particular amount of attention, since it promises to yield tools for tackling computational challenges within the NISQ~\cite{Preskill2018QuantumComputingInTheNISQEraAndBeyond,Bharti2022NoisyIntermediateScaleQuantumAlgorithms} era, where qubit numbers are limited and the lack of error correction only allows for shallow circuits.
    In a nutshell, VQAs utilize parameterized quantum circuits to reliably explore a manifold of target states and to reduce the original optimization task to optimizing the circuits' parameters.
    The parameter updates themselves are determined on a classical computer while the quantum device solely}
\newpage
\begin{itemize}
    \item[]
    \vspace*{-17pt}
\end{itemize}
serves as a sampling machine to estimate relevant expectation values.
Outsourcing the parameter updates to the classical computer typically results in shallow circuits and low demands on the coherence time, as quantum states are measured immediately after their parameterized preparation.

In addition to the shortcomings of the current quantum hardware, there are additional, conceptual hurdles to overcome.
The basic input for applying a VQA to an optimization problem is an encoding of an objective function $f$ into a multi-qubit objective Hamiltonian $H_{f}$ such that---in case of a minimization problem---optimal solutions correspond to the smallest expectation value attainable by a quantum state~\cite{Kossmann2025DeepCircuitQAOA}.
In contrast to other popular problems, like MAX-CUT or 3-Sat, the OSSP additionally demands us to not only consider the encoding of an objective function, but also the encoding of further constraints.
As a consequence, a VQA ideally has to perform its optimization only on a subspace of 'allowed/feasible' quantum states, i.e., those that respect constraints.
While in practice the presence of noise and gate infidelities prevents us from truly restricting the optimization to such a feasible subspace, the constraint-oriented quantum algorithmic design remains essential:
Feasibility-preserving gate operations, even if imperfect, can concentrate just enough amplitude within the feasible subspace to study the objective function's structure among feasible states.
{\addfontfeature{LetterSpace=1.25}This is especially important whenever the number of feasible solutions is significantly smaller than the number of all possible}
\newpage variable assignments, as is the case for the OSSP.
The major contribution of this work is a systematic analysis of symmetry structures in the OSSP, which will enable us to design a class of VQA algorithms whose logical components, by construction, restrict all parameterized states to the feasible subspace.

In general, there are two strategies for handling constrained optimization problems:
\begin{itemize}
    \item[-] \textit{Softcoded} constraints.
    Any assignment is considered feasible, but the constraints enter the objective function as additional terms and penalize assignments that were originally infeasible.
    The modified objective function results in a more complex objective Hamiltonian.
    In exchange, the ground state search can be conducted in the entire qubit space without the necessity of preserving feasibility throughout the routine.
    A comprehensible introduction and case study using Qiskit and IBM quantum computers can be found in~\cite{Sturm2023TheoryAndImplementationOfTheQuantumApproximateOptimizationAlgorithmAComprehensibleIntroductionAndCaseStudyUsingQiskitAndIBMQuantumComputers}.
    \item[-] \textit{Hardcoded} constraints.
    The objective function (and thus the objective Hamiltonian) is left unmodified, while the ground state search is restricted to the subspace corresponding to feasible solutions.
    Preservation of feasibility is typically ensured by additional gates representing the classical constraints.
\end{itemize}

Most VQAs such as the \textit{variational quantum eigensolver}~\cite{Peruzzo2014AVariationalEigenvalueSolverOnAPhotonicQuantumProcessor} or the \textit{quantum approximate optimization algorithm}~\cite{Farhi2014AQuantumApproximateOptimizationAlgorithm} are originally formulated for unconstrained problems and are thus merely applicable to instances without or softcoded constraints.
Countless use case studies have been executed on unconstrained problems like MAX-CUT~\cite{Willsch2020BenchmarkingTheQuantumApproximateOptimizationAlgorithm,Harrigan2021QuantumApproximateOptimizationOfNonPlanarGraphProblemsOnAPlanarSuperconductingProcessor} and on constrained problems with softcoded constraints, including logistic problems like the TSP~\cite{Sato2025TwoStepQuantumSearchAlgorithmForSolvingTravelingSalesmanProblems} and vehicle routing~\cite{Azad2023SolvingVehicleRoutingProblemUsingQuantumApproximateOptimizationAlgorithm,Palackal2023QuantumAssistedSolutionPathsForTheCapacitatedVehicleRoutingProblem}, as well as various scheduling problems~\cite{Tran2016AHybridQuantumClassicalApproachToSolvingSchedulingProblems,Kurowski2023ApplicationOfQuantumApproximateOptimizationAlgorithmToJobShopSchedulingProblem,Dalal2024DigitizedCounterdiabaticQuantumAlgorithmsForLogisticsScheduling}.

However, these and several other case studies indicate that softcoding the constraints often leads to suboptimal optimization landscapes or issues with feasibility~\cite{DeLaGranrive2019KnapsackProblemVariantsOfQAOAForBatteryRevenueOptimisation,Dam2021QuantumOptimizationHeuristicsWithAnApplicationToKnapsackProblems,Baker2022WassersteinSolutionQualityAndTheQuantumApproximateOptimizationAlgorithmAPortfolioOptimizationCaseStudy,Awasthi2023QuantumComputingTechniquesForMultiKnapsackProblems}.
This is why quantum algorithms with built-in possibilities for hardcoding constraints received increasing interest.
Most notably, Hadfield~\textit{et~al.}~\cite{Hadfield2019FromTheQuantumApproximateOptimizationAlgorithmToAQuantumAlternatingOperatorAnsatz} extended the quantum approximate optimization algorithm to the \textit{quantum alternating operator ansatz} (QAOA), which can also be applied to hardcoded constrained optimization problems.
Unlike its predecessor, the QAOA is formulated with a problem-dependent feasibility-preserving mixer and is therefore sensitive to different feasibility structures.
Hadfield et al. already gave explicit mixer constructions for several graph problems, the TSP, and single-machine scheduling.
Later on, suitable mixers were also constructed for various other problem classes such as the knapsack problem~\cite{Christiansen2025QuantumTreeGeneratorImprovesQAOAStateOfTheArtForTheKnapsackProblem}, capacitated vehicle routing~\cite{Xie2024AFeasibilityPreservedQuantumApproximateSolverForTheCapacitatedVehicleRoutingProblem}, and the facility locating problem~\cite{Nakada2025InductiveConstructionOfVariationalQuantumCircuitForConstrainedCombinatorialOptimization}.

Albeit there is a massive catalog~\cite{Rossi2006HandbookOfConstraintProgramming} of classical constraint analysis available for these problems, concrete mixers were mainly heuristically designed (with a few exceptions like the Grover-mixer framework~\cite{Baertschi2020GroverMixersForQAOAShiftingComplexityFromMixerDesignToStatePreparation}).
Often enough, the underlying feasibility structure was not completely exploited.
One particular interesting construction for scheduling-type problems is the \textit{constraint graph model}~\cite{Leighton1977AGraphColoringAlgorithmAGraphColoringAlgorithmForLargeSchedulingProblems,Freuder1985TakingAdvantageOfStableSetsOfVariablesInConstraintSatisfactionProblems}.
Here the bit strings are identified with vertices, joint by edges corresponding to the constraints.
This allows for investigating the feasibility structure with well-known results from graph theory.
Most notably, one can identify graph automorphisms with feasibility-preserving bit permutations.
With this work, we incorporate the entire structure into a more refined view on the QAOA and also come up with a new VQA design for OSSP instances.

In Section 2 we rigorously introduce the notion of combinatorial optimization problems, give a detailed formulation of the OSSP, cover the constraint graph model, which we will use extensively in our subsequent analysis, and describe general encoding strategies for tackling combinatorial optimization problems with the aid of quantum computers.

In Section 3 we present our theoretical and numerical results.
We first apply the constraint graph model to the general OSSP, embedding the notions of solutions and solution-preserving functions into the graph-theoretical language.
Utilizing this additional point of view, we fully characterize the group $F$ of feasibility-preserving bit permutations.
Furthermore, we uncover block structures within the OSSP-constraints.
This ultimately reveals that $F$ acts transitively on the set of all solutions.
We then draw the connection between the classical description of $F$ and specific VQA designs.
Firstly, we review how the QAOA works in general and proceed with a discussion of its main ingredients.
We give refined definitions for 'phase separator' and 'mixer' gates and detail a general construction for suitable mixers from elements of $F$.
In particular, the transitive action of $F$ is directly translated into substantial mixing properties.
Second, we introduce a new VQA suitable for OSSP instances.
It is fundamentally based on decomposing bit value permutations into products of transpositions.
In contrast to the QAOA, we can bound the number of parameters necessary to reach every possible solution:
while the number of OSSP solutions is $O(J!)$, only $O(J^{3})$ parameters are necessary.
We complement these theoretical results with two proof-of-principle numerical experiments.
The first one consists of an extensive comparison between our VQA and standard QAOA with softcoded constraints on an OSSP(2,2,4) instance.
In this noiseless simulation, we clearly observe the difficulties the standard approach has with the additional OSSP constraints;
an issue that is, by design, circumvented by our VQA.
The second experiment highlights the implementability of our VQA on today's quantum devices.
For an OSSP(1,3,3) instance, we implement our VQA on the IBM Q System One, and, additionally, on a classical noisy simulator.

\section{\label{section:MaterialsAndMethods}Material and methods}

For the readers' convenience we will review the basic notion of combinatorial optimization problems (COPs), especially the open-shop scheduling problem.
We also briefly introduce the constraint graph model and cover the very basics of problem encoding onto quantum computers.
Throughout this work, we will use the shorthand $[N] \coloneqq \{1, \ldots, N\}$ where $N$ is any natural number.

\subsection{\label{subsection:ConstrainedCombinatorialOptimization}Constrained combinatorial optimization}

In the following, we restrict to minimization problems, as maximization tasks may be considered analogously.
Then a generic COP of \textit{size} $N$ with $A$ constraints is of the form
\begin{align}\label{equation:COP}
    \min_{\bm{z} \scriptin \bits} f(\bm{z})\quad \text{s.t.}\ c_{a}(\bm{z}) = 1,\quad a \in [A],
\end{align}
where
\begin{itemize}
    \item[-] $\bits \coloneqq \{0, 1\}^{N}$ denotes the set of all bit strings of length $N$;
    \item[-] $f : \bits \rightarrow \R$ is the \textit{objective function} to be minimized;
    \item[-] $c_{a} : \bits \rightarrow \{0, 1\}$ are the \textit{constraints}.
    \vspace{-6pt}
\end{itemize}
Accordingly, a bit string $\bm{z}$ is said to fulfill a constraint $c_{a}$ if $c_{a}(\bm{z}) = 1$.
In the following, we will refer to COPs formally as triples $(N, f, \{c_{a}\})$.
For a given COP $\COP$, we define its \textit{solution set} as the set of all bit strings, fulfilling every constraint:
\vspace{-6pt}
\begin{align}
    S(\COP) \coloneqq \{\bm{z} \in \bits \defcolon c_{a}(\bm{z}) = 1,\quad a \in [A]\}.
\end{align}
Furthermore, the \textit{optimal solution set} is the set of all solution bit strings minimizing $f$, i.e.,
\vspace{-6pt}
\begin{align}
    S_{\min}(\COP) \coloneqq \left\{\bm{z} \in S(\COP) \defcolon f(\bm{z}) = \min_{\bm{z}' \scriptin S(\COP)} f(\bm{z}')\right\} \subseteq S(\COP).
\end{align}

We now examine particular types of constraints:
For two bit strings $\bm{z}, \bm{z}' \in z(N)$ the bit-wise \textit{and} operation produces a new bit string $\bm{z} \wedge \bm{z}' \in z(N)$.
Let $\abs{\bm{z}}$ denote the \textit{Hamming weight} of $\bm{z}$.
For an index subset $I \subseteq [N]$, we define the bit string $\bm{z}_{I} \in z(N)$ to fulfill $z_{n} = 1$ if $n \in I$.
The \textit{one-hot constraint} and the \textit{at-most-one constraint} associated with the index set $I$ are
\vspace{-6pt}
\begin{align}
    \zeta_{I}(\bm{z}) &\coloneqq \begin{cases}
        1,& \text{if } \abs{\bm{z} \wedge \bm{z}_{I}} = 1\\
        0,& \text{otherwise}
    \end{cases}\quad\text{and}\\
    \eta_{I}(\bm{z}) &\coloneqq \begin{cases}
        1,& \text{if } \abs{\bm{z} \wedge \bm{z}_{I}} \leq 1\\
        0,& \text{otherwise.}
    \end{cases}
\end{align}

\subsection{\label{subsection:Open-ShopScheduling}Open-shop scheduling}

Let us formally introduce the OSSP, depending on the three parameters
\vspace{-6pt}
\begin{itemize}
    \item[-] $M$ is the number of machines;
    \item[-] $T$ is the number of time slots per machine;
    \item[-] $J$ is the number of jobs.
\end{itemize}
The $J$ jobs should be distributed to the $M T$ available positions such that
\vspace{-6pt}
\begin{itemize}
    \item[$\mathbf{J}$] represents every job gets performed precisely once;
    \item[$\mathbf{P}$] indicates that no position is filled with more than one job.
\end{itemize}
Note that OSSP($1, T, T$) has the exact same structure as the $T$-city TSP.

In order to bring OSSP\,$\coloneqq$\,OSSP($M, T, J$) into the form \textbf{(\ref{equation:COP})}, we introduce $N = M T J$ bits, identify $[N] \cong [M] \times [T] \times [J]$, and set
\begin{align}
    z_{m t j} \coloneqq \begin{cases}
        1,& \text{if job }j\text{ runs on machine }m\text{ at time }t,\\
        0,& \text{otherwise.}
    \end{cases}
\end{align}
The \textit{job assignment constraints} $\mathbf{J}$ and the \textit{position assignment constraints} $\mathbf{P}$ then read
\begin{align}
    &\mathbf{J}:\ \sum_{m = 1}^{M} \sum_{t = 1}^{T} z_{m t j} = 1,\quad j \in [J]\quad \text{and} \\
    &\mathbf{P}:\ \sum_{j = 1}^{J} z_{m t j} \leq 1,\quad (m, t) \in [M] \times [T].
\end{align}
Thus $\mathbf{J}$ and $\mathbf{P}$ are one-hot and at-most-one constraints, respectively.
Namely, we have
\begin{align}
    &\mathbf{J}: c_{j} \coloneqq \zeta_{\Delta_{(j)}},\ j \in [J]\quad \text{and} \\
    &\mathbf{P}: c_{m, t} \coloneqq \eta_{\Delta_{(m, t)}},\ (m, t) \in [M] \times [T]
\end{align}
with \textit{job blocks} $\Delta_{(j)} \coloneqq [M] \times [T] \times \{j\}$ and \textit{position blocks} $\Delta_{(m, t)} \coloneqq \{m\} \times \{t\} \times [J]$.
The constraints are equivalently captured in the coordinate relation
\vspace{-6pt}
\begin{align}\label{equation:OSSPRelation}
    (m, t, j) \sim (m', t', j') :\Longleftrightarrow\,(m = m' \ \land \ t = t')\ \vee\ j = j' .
\end{align}
This means that if $(m, t, j) \sim (m', t', j')$ then any feasible solution $\bm{z}$ cannot have both bits $z_{m t j}$ and $z_{m' t' j'}$ set to one.
Note that `$\sim$' is reflexive and symmetric, but not transitive.
The OSSP solution set (expressed as a coordinate set) is explicitly given by
\begin{align}\label{equation:OSSPSolutionSet}
    \bigcup&\Big\{\!\{(m_{j} , t_{j}, j)\! :\! j \in [J]\}\! :\! (m_{1}, t_{1})\! \neq \! \cdots \! \neq \! (m_{J}, t_{J})\! \in \! [M] \! \times \! [T]\Big\}.
\end{align}
Therefore, it possesses $(M T)! / (M T - J)!$ solutions.

Very common objectives of scheduling tasks are the minimization of the schedule's makespan or of the total completion time.
For our formulation of the problem where jobs do not have to run on all machines and each job's completion takes up exactly one time slot, these two objective are, however, not of much interest as equally balancing the assignment of jobs over all machines at the earliest time slots readily minimizes both quantities.
Instead, we introduce machine- and time-sensitive execution costs for each job, $\omega_{m t j} \in \R$, and impose the objective of minimizing the schedule's cumulative costs:
\vspace{-3pt}
\begin{align}\label{equation:ObjectiveFunction}
    f : z(N) \rightarrow \R,\quad \bm{z} \mapsto \sum_{m = 1}^{M} \sum_{t = 1}^{T} \sum_{j = 1}^{J} \omega_{m t j} z_{m t j} \eqqcolon \bm{\omega} \cdot \bm{z},
\end{align}
where $\bm{\omega} \in \R^{N}$ represents the cost vector.
Therefore, the objective function is linear and reduces to evaluating the inner product of $\bm{\omega}$ with the bit string argument.

\subsection{Constraint graph model}\label{subsection:ConstraintGraphModel}

The solution set of a COP $\COP = (N, f, \{c_{a}\})$ can be further studied by introducing the so-called \textit{constraint graph}~\cite{Leighton1977AGraphColoringAlgorithmAGraphColoringAlgorithmForLargeSchedulingProblems,Freuder1985TakingAdvantageOfStableSetsOfVariablesInConstraintSatisfactionProblems}.
First we identify $[N]$ with a vertex set $V = \{v_{1}, \ldots, v_{N}\}$.
Moreover, the bit string $\bm{z}_{I}$, associated to the subset $I \subseteq [N]$, is identified with the subset of vertices $V_{I} \coloneqq \{v_{n} \defcolon n \in I\} \subseteq V$.

The constraints enter as edges of the graph:
Two vertices $v_{m}$ and $v_{n}$ are joint by an edge if there is a constraint that prohibits the two bits $z_{m}$ and $z_{n}$ from taking the value 1 at the same time.
Since this construction is symmetric, the constraint graph $G = (V, E)$ is undirected.
Thus, the edges are unordered pairs of vertices and we may address an edge $e \in E$ with its end points.
Assuming that we have defined a coordinate relation '$\sim$' similar to \textbf{(\ref{equation:OSSPRelation})}, we can express the set of edges simply as $E = \{(v_{m}, v_{n}) \defcolon m \sim n\}$.

For one-hot and at-most-one constraints we can be more concrete:
If $\zeta_{I} \in \{c_{a}\}$ or $\eta_{I} \in \{c_{a}\}$ all vertices $v_{n}$ with $n \in I$ are mutually connected, i.e., they form a clique in the constraint graph.
\textbf{\autoref{figure:OSSP224Graph}} shows the constraint graph of an OSSP$(2, 2, 4)$-instance.

A solution to $\COP$ does not violate any constraint.
Thus it corresponds to an independent set (or coclique) in the associated constraint graph $G(\COP)$ (see \autoref{figure:OSSP224GraphSolution} for such a graphical solution to the OSSP$(2, 2, 4)$ instance).
If $\COP$ incorporates $J$ one-hot constraints and some additional at-most-one constraints (such as the OSSP), every solution has Hamming weight $J$.
Accordingly, we call any vertex subset $W \subseteq V$ a \textit{solution} to $\COP$ if
\begin{itemize}
    \item[(i)] $\abs{W} = J$;
    \item[(ii)] $W$ is an independent set.
\end{itemize}

Moreover, we say that a permutation $\rho : V \rightarrow V$ \textit{preserves feasibility} if $\rho(W)$ is again a solution whenever $W \subseteq V$ is a solution.
We denote with $F$ the set of all feasibility-preserving permutations.
As the composition of two feasibility-preserving permutations preserves feasibility again and the identity also preserves feasibility, one readily deduces that $F$ is a subgroup of $\Sym(V)$.
Via the identification of bit strings with vertex subsets, $F$ can also be interpreted as acting on the solution set $S(\COP)$, i.e., on a set of bit strings.
However, the notions of solutions and of $F$ can also be considered independently of any underlying COP.

We remark that we could alternatively have declared the complement graph of $G(\COP)$ as the actual constraint graph.
Then, solutions would correspond to cliques instead of cocliques, but the underlying structure would not change since a graph and its complement have the same automorphism group.
This equivalent point of view is noteworthy since many results in graph theory are formulated in terms of cliques.
However, we proceed with our original definition.

\begin{figure*}
    \centering
    \begin{tikzpicture}[scale=0.95]
        \tikzset{
            myarrow/.style={
                ->,
                >=latex',
                thick
            },
            mycircle/.style={
                shape=circle,
                draw=green,
                inner sep=0pt,
                text width=6.5mm,
                align=center,
            }
        }
        \node (A1) at (0,0) [mycircle] {\scriptsize 1,1,1};
        \node (A2) at (1.5,0) [mycircle] {\scriptsize 1,2,1};
        \node (A3) at (4.5,0) [mycircle] {\scriptsize 1,1,2};
        \node (A4) at (6,0) [mycircle] {\scriptsize 1,2,2};
        \node (A5) at (9,0) [mycircle] {\scriptsize 1,1,3};
        \node (A6) at (10.5,0) [mycircle] {\scriptsize 1,2,3};
        \node (A7) at (13.5,0) [mycircle] {\scriptsize 1,1,4};
        \node (A8) at (15,0) [mycircle] {\scriptsize 1,2,4};

        \node (B1) at (0,-1.5) [mycircle] {\scriptsize 2,1,1};
        \node (B2) at (1.5,-1.5) [mycircle] {\scriptsize 2,2,1};
        \node (B3) at (4.5,-1.5) [mycircle] {\scriptsize 2,1,2};
        \node (B4) at (6,-1.5) [mycircle] {\scriptsize 2,2,2};
        \node (B5) at (9,-1.5) [mycircle] {\scriptsize 2,1,3};
        \node (B6) at (10.5,-1.5) [mycircle] {\scriptsize 2,2,3};
        \node (B7) at (13.5,-1.5) [mycircle] {\scriptsize 1,1,4};
        \node (B8) at (15,-1.5) [mycircle] {\scriptsize 1,2,4};

        \draw[bend left=19] (A1) to (A3);
        \draw[bend left=17] (A1) to (A5);
        \draw[bend left=15] (A1) to (A7);
        \draw[bend left=19] (A2) to (A4);
        \draw[bend left=17] (A2) to (A6);
        \draw[bend left=15] (A2) to (A8);
        \draw[bend left=19] (A3) to (A5);
        \draw[bend left=17] (A3) to (A7);
        \draw[bend left=19] (A4) to (A6);
        \draw[bend left=17] (A4) to (A8);
        \draw[bend left=19] (A6) to (A8);

        \draw[bend right=19] (B1) to (B3);
        \draw[bend right=17] (B1) to (B5);
        \draw[bend right=15] (B1) to (B7);
        \draw[bend right=19] (B2) to (B4);
        \draw[bend right=17] (B2) to (B6);
        \draw[bend right=15] (B2) to (B8);
        \draw[bend right=19] (B3) to (B5);
        \draw[bend right=17] (B3) to (B7);
        \draw[bend right=19] (B4) to (B6);
        \draw[bend right=17] (B4) to (B8);
        \draw[bend right=19] (B6) to (B8);

        \draw (A1) to (A2);
        \draw (A1) to (B1);
        \draw (A1) to (B2);
        \draw (A2) to (B1);
        \draw (A2) to (B2);
        \draw (B1) to (B2);

        \draw (A3) to (A4);
        \draw (A3) to (B3);
        \draw (A3) to (B4);
        \draw (A4) to (B3);
        \draw (A4) to (B4);
        \draw (B3) to (B4);

        \draw (A5) to (A6);
        \draw (A5) to (B5);
        \draw (A5) to (B6);
        \draw (A6) to (B5);
        \draw (A6) to (B6);
        \draw (B5) to (B6);

        \draw (A7) to (A8);
        \draw (A7) to (B7);
        \draw (A7) to (B8);
        \draw (A8) to (B7);
        \draw (A8) to (B8);
        \draw (B7) to (B8);

        \node (origin) at (-0.6,0.7) {};
        \draw[myarrow] (origin.south) -- ++(1,0) node[midway, sloped, above] {$t$};
        \draw[myarrow] (origin.south) -- ++(0,-1) node[rotate=180, midway, sloped, above] {$m$};

        \node (C1) at (0,-4.5) [mycircle] {1};
        \node (C2) at (1.5,-4.5) [mycircle] {5};
        \node (C3) at (4.5,-4.5) [mycircle] {2};
        \node (C4) at (6,-4.5) [mycircle] {6};
        \node (C5) at (9,-4.5) [mycircle] {3};
        \node (C6) at (10.5,-4.5) [mycircle] {7};
        \node (C7) at (13.5,-4.5) [mycircle] {4};
        \node (C8) at (15,-4.5) [mycircle] {8};

        \node (D1) at (0,-6) [mycircle] {7};
        \node (D2) at (1.5,-6) [mycircle] {11};
        \node (D3) at (4.5,-6) [mycircle] {8};
        \node (D4) at (6,-6) [mycircle] {12};
        \node (D5) at (9,-6) [mycircle] {9};
        \node (D6) at (10.5,-6) [mycircle] {13};
        \node (D7) at (13.5,-6) [mycircle] {10};
        \node (D8) at (15,-6) [mycircle] {14};

        \draw[bend left=19] (C1) to (C3);
        \draw[bend left=17] (C1) to (C5);
        \draw[bend left=15] (C1) to (C7);
        \draw[bend left=19] (C2) to (C4);
        \draw[bend left=17] (C2) to (C6);
        \draw[bend left=15] (C2) to (C8);
        \draw[bend left=19] (C3) to (C5);
        \draw[bend left=17] (C3) to (C7);
        \draw[bend left=19] (C4) to (C6);
        \draw[bend left=17] (C4) to (C8);
        \draw[bend left=19] (C6) to (C8);

        \draw[bend right=19] (D1) to (D3);
        \draw[bend right=17] (D1) to (D5);
        \draw[bend right=15] (D1) to (D7);
        \draw[bend right=19] (D2) to (D4);
        \draw[bend right=17] (D2) to (D6);
        \draw[bend right=15] (D2) to (D8);
        \draw[bend right=19] (D3) to (D5);
        \draw[bend right=17] (D3) to (D7);
        \draw[bend right=19] (D4) to (D6);
        \draw[bend right=17] (D4) to (D8);
        \draw[bend right=19] (D6) to (D8);

        \draw (C1) to (C2);
        \draw (C1) to (D1);
        \draw (C1) to (D2);
        \draw (C2) to (D1);
        \draw (C2) to (D2);
        \draw (D1) to (D2);

        \draw (C3) to (C4);
        \draw (C3) to (D3);
        \draw (C3) to (D4);
        \draw (C4) to (D3);
        \draw (C4) to (D4);
        \draw (D3) to (D4);

        \draw (C5) to (C6);
        \draw (C5) to (D5);
        \draw (C5) to (D6);
        \draw (C6) to (D5);
        \draw (C6) to (D6);
        \draw (D5) to (D6);

        \draw (C7) to (C8);
        \draw (C7) to (D7);
        \draw (C7) to (D8);
        \draw (C8) to (D7);
        \draw (C8) to (D8);
        \draw (D7) to (D8);

        \draw [dashed] (0.75,-5.25) circle (1.5cm);
        \draw [dashed] (5.25,-5.25) circle (1.5cm);
        \draw [dashed] (9.75,-5.25) circle (1.5cm);
        \draw [dashed] (14.25,-5.25) circle (1.5cm);

        \node (jb1) at (0.9,-7.6) {$\Delta_{(1)}$};
        \node (jb2) at (5.4,-7.6) {$\Delta_{(2)}$};
        \node (jb3) at (9.9,-7.6) {$\Delta_{(3)}$};
        \node (jb4) at (14.4,-7.6) {$\Delta_{(4)}$};
    \end{tikzpicture}
    \caption{OSSP(2,2,4) constraint graph.
        In the above graph, the vertices are labeled using the coordinate system $(m, t, j)$.
        In addition, the three job blocks $\Delta_{(j)}$, $j \in [4]$ are depicted.}\label{figure:OSSP224Graph}
\end{figure*}

\subsection{Encoding on quantum computers}\label{subsection:EncodingOnQuantumComputers}

In the following, we recall the standard encoding procedure for addressing a general COP $\COP = (N, f, \{c_{a}\})$ with the aid of a quantum computer.
All relevant quantum computing essentials are covered in~\cite{Nielsen2010QuantumComputationAndQuantumInformation}.
First, we identify each bit string $\bm{z}$ with a computational basis state $\ket{\bm{z}}$ of the $N$-qubit space $\hil \coloneqq \C^{2^{N}}$.
This induces a representation of functions over $\bits$ as linear operators on $\hil$.
Namely, the classical objective function $f$ is mapped to an objective Hamiltonian, diagonal in the computational basis
\begin{align}
    f \mapsto C \coloneqq \sum_{\bm{z} \in \bits} f(\bm{z}) \ketbra{\bm{z}}.
\end{align}
For unconstrained problems the minimization task is then equivalent to finding a computational basis state, which is a ground state of $C$.
However, in the constrained case, the ground state search has to be restricted to the \textit{solution space}
\begin{align}
    \mathcal{S} \coloneqq \spn\{\ket{\bm{z}} \defcolon \bm{z} \in S(\COP)\} \subsetneq \hil.
\end{align}
The \textit{optimal solution space}
\begin{align}
    \mathcal{S}_{\min} \coloneqq \spn\{\ket{\bm{z}} \defcolon \bm{z} \in S_{\min}(\COP)\}
\end{align}
then is a subspace of $\mathcal{S}$ and is the eigenspace of $C\vert_{\mathcal{S}}$, corresponding to its smallest eigenvalue.
Unlike $\hil$, $\mathcal{S}$ does not admit any favorable tensor product structure in general.
This is precisely what makes constrained optimization more challenging on a quantum computer.

\section{\label{section:Results}Results}

In this section we first investigate the constraint graph of an OSSP instance.
Our focus lies on determining the feasibility-preserving subgroup $F$ and its properties.
For brevity, we will not distinguish between the COP and its constraint graph in what follows.

Subsequently, we elaborate on the QAOA method for tackling combinatorial optimization with the aid of quantum computers.
Here, we connect our notion of feasibility-preserving groups to desirable properties of QAOA circuits.
These theoretical considerations cumulate in the formulation of our new VQA, specifically tailored to the OSSP, with its guaranteed upper bound on variational parameters required to reach the optimal solution.

Finally, as a proof of principle, we numerically test our proposed VQA on a 16-qubit OSSP instance in a noiseless simulation as well as on a 9-qubit instance on the IBM Q System One, i.e.,\ on real quantum hardware.
The latter experiment is accompanied by a classical simulation with a noise model.

\subsection{\label{subsection:Feasbility-PreservingGraphAutomorphisms}Feasibility-preserving graph automorphisms}

Let $G = (V, E)$ be a graph.
Recall that a graph automorphism is a bijection $\varphi : V \rightarrow V$ such that $\varphi$ as well as $\varphi^{-1}$ preserve adjacency.
One can actually prove that any bijective graph homomorphism between $G$ and itself is already a graph automorphism.
We start with the general observation that graph automorphisms are always feasibility-preserving.

\begin{Proposition}[\cite{Binkowski2022ConstraintGraphModelAnalysisOfTheQuantumAlternatingOperatorAnsatz}]\label{proposition:GraphAutomorphismsPreserveFeasibility}
    Let $G = (V, E)$ be a graph with solution set $S$.
    It holds that $\Aut(G) \subseteq F$.
\end{Proposition}

\begin{proof}
    Let $\varphi \in \Aut(G)$ and let $W \in S$ be arbitrary.
    Since $\varphi : V \rightarrow V$ is bijective, it holds that $\abs{\varphi(W)} = \abs{W} = J$.
    Let $\varphi(v), \varphi(w) \in \varphi(W)$.
    {\addfontfeature{LetterSpace=-2.0}Since $W$ is an independent set, it holds that $v w \notin E$.}
    {\addfontfeature{LetterSpace=-2.0}With the isomorphism property of $\varphi^{-1}$, it follows that $\varphi(v) \varphi(w) \notin E$;}
    {\addfontfeature{LetterSpace=-2.0}hence $\varphi(W)$ is again an independent set.}
    {\addfontfeature{LetterSpace=-2.0}This shows that $\Aut(G) \subseteq F$.}
\end{proof}

For general graphs $G$, $F$ will be a strict superset of $\Aut(G)$.
In case of the OSSP graph, however, equality holds.
In order to prove this we utilize a simple auxiliary result.

\begin{Proposition}[\cite{Binkowski2022ConstraintGraphModelAnalysisOfTheQuantumAlternatingOperatorAnsatz}]\label{proposition:Converse}
    Let $G = (V, E)$ be a graph with solution set $S$.
    If for all non-adjacent $v, w \in V$ there exists $W \in S$ such that $v, w \in W$, then $\Aut(G) = F$.
\end{Proposition}

\begin{proof}
    By Proposition \autoref{proposition:GraphAutomorphismsPreserveFeasibility} it suffices to show that $F \subseteq \Aut(G)$.
    Let $\rho \in F$ and $v, w \in V$ with $\rho(v) \rho(w) \notin E$.
    Then, there exists $W \in S$ so that $\rho(v), \rho(w) \in W$.
    Since $F$ is a group, $\rho^{-1}$ is also feasibility-preserving, hence $v, w \in \rho^{-1}(W) \in S$ are non-adjacent.
    Thus the bijective map $\rho$ fulfills $(\rho(v) \rho(w) \notin E \Rightarrow v w \notin E)$, which is an equivalent characterization of a graph automorphism.
\end{proof}

Consider again the OSSP solution set \textbf{(\ref{equation:OSSPSolutionSet})}.
Note that any pair $(m, t, j)$, $(m', t', j')$ corresponding to non-adjacent vertices, i.e., $(m, t) \neq (m', t')$ and $j \neq j'$, can be augmented to form a solution to the OSSP.
Therefore, we can apply \autoref{proposition:Converse} and conclude that \mbox{$\Aut(\OSSP) = F$}.

\subsection{\label{subsection:DeterminingF}Determining \texorpdfstring{$\mathbf{\emph{F}}$}{F}}

Since the relation \textbf{(\ref{equation:OSSPRelation})} is divided in two logically disjoint parts, one concludes that the constraint graph is given by the Cartesian product~\cite{Ilmrich2000ProductGraphs} of the two graphs $K^{P}$ and $K^{J}$ for $P \coloneqq M T$, where $K^{n}$ is the complete graph with $n$ vertices.
With some effort one can show the following theorem which we owe to personal correspondence with Benjamin Sambale.
\begin{Theorem}\label{theorem:componentsautogroup}
    Let $G$ be a graph and $H_{1}, \ldots, H_{k}$ be representatives of the isomorphism types of indecomposable components~\cite{Ilmrich2000ProductGraphs} of $G$. \\
    Let $m_{i}$ be the number of indecomposable components of $G$ that are isomorphic to $H_{i}$.
    Then, it holds
    \begin{align}
        \Aut(G) \cong \bigtimes_{i = 1}^{k} \big(\underbrace{[\Aut(H_{i}) \times \cdots \times \Aut(H_{i})]}_{m_{i}\text{-}\Times} \rtimes S_{m_{i}}\big).
    \end{align}
\end{Theorem}
\autoref{theorem:componentsautogroup} states that knowing the indecomposable components of a graph reduces the problem of determining its automorphism group to the symmetries of the indecomposable components.
In the situation of the OSSP we are mostly done since it is not difficult to show that complete graphs are indecomposable.
The automorphism group of complete graphs simply is the whole symmetric group over its vertex set because all vertex permutations are graph automorphisms.
Therefore, the automorphism group of the OSSP is given by
\begin{align}
    F = \Aut(\OSSP) \cong \begin{cases}
        (S_{J} \times S_{J}) \rtimes S_{2} & P = J,\\
        \hphantom{(}S_{P} \times S_{J} & P > J.
    \end{cases}
\end{align}

We observe that the \textit{busy} case ($P = J$) differs structurally from the \textit{non-busy} case ($P > J$).
For the latter case we simply obtain an automorphism group that is the direct product of \textit{position permutations} and \textit{job permutations}.
However, in the busy case, we obtain a \textit{wreath product}~\cite{Roman1995AnIntroductionToTheTheoryOfGroups} structure.
There is an additional non-trivial automorphism that interchanges between position and jobs.
In our subsequent analysis, however, we will consider the subgroup
\begin{align}
    F' \coloneqq S_{J} \times S_{J} \leq (S_{J} \times S_{J}) \rtimes S_{2}
\end{align}
in order to treat both cases similarly.
The concrete action of an element $(\sigma, \tau) \in S_{P} \times S_{J}$ on a coordinate tuple $(m, t, j)$ is given by
\begin{align}
    {}^{(\sigma, \tau)} (m, t, j) = (\sigma(m, t), \tau(j)).
\end{align}

We now argue that $S_{P} \times \{\id\} \leq F$ acts transitively on the solution set \textbf{(\ref{equation:OSSPSolutionSet})}.
Let $s, s' \subset [M] \times [T] \times [J]$ be two solutions, hence there exist $m_{1}, \ldots, m_{J}, m_{1}', \ldots, m_{J}' \in [M]$ and $t_{1}, \ldots, t_{J}, t_{1}', \ldots, t_{J}' \in [T]$ such that
\begin{align}
    s &= \{(m_{1}, t_{1}, 1), \ldots, (m_{J}, t_{J}, J)\}\text{ and}\\
    s' &= \{(m_{1}', t_{1}', 1), \ldots, (m_{J}', t_{J}', J)\}.
\end{align}
Since $S_{P}$ acts $P$-transitively on $[M] \times [T]$ and $P \geq J$, there exists $\sigma \in S_{P}$ such that
\begin{align}
    {}^{\sigma}(m_{j}, t_{j}) = (m_{j}', t_{j}')
\end{align}
holds for all $j \in [J]$, i.e., ${}^{(\sigma, \id)}s = s'$.
Thus we have just concluded
\begin{Theorem}[\cite{Binkowski2022ConstraintGraphModelAnalysisOfTheQuantumAlternatingOperatorAnsatz,Kossmann2023AQuantumAlgorithmWithGroupTheory}]\label{theorem:transitivity}
    The action of $F$ on the solution set $S$ is transitive for the OSSP.
\end{Theorem}

\subsection{\label{subsection:BlockStructure}Block structure}

We further characterize the group action of $F$ (resp.\ $F'$) via block systems.
Given a group $G$ acting on some set $X$, a subset $\Delta \subset X$ with $1 < \abs{\Delta} < \abs{X}$ is called a \textit{block}~\cite{Sambale2017EndlichePermutationsgruppen} of $G$ if for all $g \in G$ it holds that ${}^{g} \Delta = \Delta \ \vee \ {}^g \Delta \cap \Delta  = \emptyset$.
A partition of $X$ into blocks of $G$ is then called a \textit{block system}.

It is immediately clear that the collection of job blocks $\Delta_{(j)}$ and of position blocks $\Delta_{(m, t)}$ each form a block system of $F$ (resp.\ $F'$).
Furthermore, one readily verifies that $F$ (resp.\ $F'$) acts transitively on $[M] \times [T] \times [J]$.
Recall that the job and position blocks result from one-hot and at-most-one constraints and are therefore cliques in the constraint graph.
That is, all vertices in a block are adjacent, which implies that a solution is a subset $s \subset [M] \times [T] \times [J]$ such that each element in $s$ belongs to exactly one block in each of the two partitions into position blocks and job blocks.
In both cases this yields a bijection between elements in $s$ and the job blocks.
In the busy case there is an additional bijection between elements in $s$ and the position block while in the non-busy case $s$ only occupies a subset of all position blocks.
\begin{figure*}[!hb]
    \centering
    \begin{tikzpicture}[scale=0.97]
        \tikzset{
            myarrow/.style={
                ->,
                >=latex',
                thick
            },
            mycircle/.style={
                shape=circle,
                draw=green,
                inner sep=0pt,
                text width=6.5mm,
                align=center,
            },
            solcircle/.style={
                fill=blue,
                shape=circle,
                draw=green,
                inner sep=0pt,
                text width=6.5mm,
                align=center,
            }
        }
        \node (A1) at (0,0) [solcircle] {\scriptsize 1,1,1};
        \node (A2) at (1.5,0) [mycircle] {\scriptsize 1,2,1};
        \node (A3) at (4.5,0) [mycircle] {\scriptsize 1,1,2};
        \node (A4) at (6,0) [mycircle] {\scriptsize 1,2,2};
        \node (A5) at (9,0) [mycircle] {\scriptsize 1,1,3};
        \node (A6) at (10.5,0) [mycircle] {\scriptsize 1,2,3};
        \node (A7) at (13.5,0) [mycircle] {\scriptsize 1,1,4};
        \node (A8) at (15,0) [solcircle] {\scriptsize 1,2,4};

        \node (B1) at (0,-1.5) [mycircle] {\scriptsize 2,1,1};
        \node (B2) at (1.5,-1.5) [mycircle] {\scriptsize 2,2,1};
        \node (B3) at (4.5,-1.5) [mycircle] {\scriptsize 2,1,2};
        \node (B4) at (6,-1.5) [solcircle] {\scriptsize 2,2,2};
        \node (B5) at (9,-1.5) [solcircle] {\scriptsize 2,1,3};
        \node (B6) at (10.5,-1.5) [mycircle] {\scriptsize 2,2,3};
        \node (B7) at (13.5,-1.5) [mycircle] {\scriptsize 1,1,4};
        \node (B8) at (15,-1.5) [mycircle] {\scriptsize 1,2,4};

        \draw[bend left=19] (A1) to (A3);
        \draw[bend left=17] (A1) to (A5);
        \draw[bend left=15] (A1) to (A7);
        \draw[bend left=19] (A2) to (A4);
        \draw[bend left=17] (A2) to (A6);
        \draw[bend left=15] (A2) to (A8);
        \draw[bend left=19] (A3) to (A5);
        \draw[bend left=17] (A3) to (A7);
        \draw[bend left=19] (A4) to (A6);
        \draw[bend left=17] (A4) to (A8);
        \draw[bend left=19] (A6) to (A8);

        \draw[bend right=19] (B1) to (B3);
        \draw[bend right=17] (B1) to (B5);
        \draw[bend right=15] (B1) to (B7);
        \draw[bend right=19] (B2) to (B4);
        \draw[bend right=17] (B2) to (B6);
        \draw[bend right=15] (B2) to (B8);
        \draw[bend right=19] (B3) to (B5);
        \draw[bend right=17] (B3) to (B7);
        \draw[bend right=19] (B4) to (B6);
        \draw[bend right=17] (B4) to (B8);
        \draw[bend right=19] (B6) to (B8);

        \draw (A1) to (A2);
        \draw (A1) to (B1);
        \draw (A1) to (B2);
        \draw (A2) to (B1);
        \draw (A2) to (B2);
        \draw (B1) to (B2);

        \draw (A3) to (A4);
        \draw (A3) to (B3);
        \draw (A3) to (B4);
        \draw (A4) to (B3);
        \draw (A4) to (B4);
        \draw (B3) to (B4);

        \draw (A5) to (A6);
        \draw (A5) to (B5);
        \draw (A5) to (B6);
        \draw (A6) to (B5);
        \draw (A6) to (B6);
        \draw (B5) to (B6);

        \draw (A7) to (A8);
        \draw (A7) to (B7);
        \draw (A7) to (B8);
        \draw (A8) to (B7);
        \draw (A8) to (B8);
        \draw (B7) to (B8);

        \node (origin) at (-0.6,0.7) {};
        \draw[myarrow] (origin.south) -- ++(1,0) node[midway, sloped, above] {$t$};
        \draw[myarrow] (origin.south) -- ++(0,-1) node[rotate=180, midway, sloped, above] {$m$};
    \end{tikzpicture}
    \caption{OSSP(2,2,4) constraint graph with feasible solution.
        The colored vertices are a maximally independent set and thus constitute a feasible solution.}\label{figure:OSSP224GraphSolution}
\end{figure*}
Since each element of a solution exactly corresponds to one position and to one job block, we can capture the action of $F$ (resp.\ $F'$) on the solution set $S$ equivalently as its action on the blocks.
Here, \cite[Proposition 1.37]{Sambale2017EndlichePermutationsgruppen} states that if the blocks are maximal with respect to inclusion (which they are here), then the block-wise action of $F$ is primitive, i.e., it does not possess blocks on its own.

We lastly focus on the busy case.
Since the normal action of $S_{J}$ on $J$ elements is \textit{sharply transitive}, for every two blocks there is exactly one element in $S_{J}$ that maps between them.
The solutions are now precisely the $J!$ permutations of these blocks.
Furthermore, each of the two copies of $S_{J}$ is a stabilizer for one of the block structures and is also a normal subgroup of $F'$.
Therefore we can identify the solutions with one of those subgroups and obtain a \textit{regular}~\cite{Sambale2017EndlichePermutationsgruppen} group action.
Note that for the non-busy case the unique identification of solutions with elements of $S_{P}$ or $S_{J}$ is not possible as one has
\begin{align}
    \abs{S_{J}} = J! < \frac{P!}{(P - J)!} < P! = \abs{S_{P}}.
\end{align}
We leave it as an open problem to find and characterize subgroups of $S_{P}$ that are in bijection with the solution set.

\subsection{Relation to the QAOA}\label{subsection:RelationToTheQAOA}

We now utilize our knowledge about feasibility-preserving permutations to study and design VQAs.
The underlying connection is due to the fact that classical operations on bit strings may be considered as quantum permutation operators acting on the associated qubit space $\hil$:
Consider a COP $\COP$.
Any group $G$ that acts on $\bits$ (resp.\ on $S(\COP)$) also acts on the computational basis (resp.\ on the feasible computational basis states) via $^{g}\ket{\bm{z}} \coloneqq \ket{^{g}\bm{z}}$, $g \in G$.
By linearity, we can extend this action to the whole space $\hil$ (resp.\ $\mathcal{S}$), yielding linear operators $\rho(g) \in \lo(\hil)$ (resp.\ $\rho_{\mathcal{S}}(g) \in \lo(\mathcal{S})$), which are simply permutation matrices in the computational basis.
In group representation theory this construction is known as the \textit{permutation representation}~\cite{Sagan2001TheSymmetricGroup}.

We start by connecting our group-theoretic picture for mixing operations with the established QAOA framework.
Consider a COP with encoded objective Hamiltonian $C$, solution space $\mathcal{S} \subseteq \hil$, and optimal solution space $\mathcal{S}_{\min} \subseteq \mathcal{S}$.
Starting from a \textit{feasible} initial state $\ket{\iota} \in \mathcal{S}$, the QAOA prescribes to alternately apply parameterized \textit{phase separator} gates $\phaseu(\gamma)$ and \textit{mixer} gates $\mixu(\beta)$ $p$ times, where $p$ is the circuit depth.
This yields a parameterized trial state
\begin{align}
    \ket{\vec{\beta}, \vec{\gamma}} \coloneqq \left(\prod_{o = 1}^{p} \mixu(\beta_{o}) \phaseu(\gamma_{o})\right) \ket{\iota}.
\end{align}
After preparing $\ket{\vec{\beta}, \vec{\gamma}}$, one can now estimate
\begin{align}
    F_{p}(\vec{\beta}, \vec{\gamma}) \coloneqq \braket{\vec{\beta}, \vec{\gamma} | C | \vec{\beta}, \vec{\gamma}}
\end{align}
via repeated sampling on the quantum computer and pass this quantity to a classical optimizer, which updates the parameters $\vec{\beta}$, $\vec{\gamma}$ in order to minimize $F_{p}(\vec{\beta},\vec{\gamma})$.
On real quantum hardware, the required finite coherence time as well as infidelities of the parameterized quantum circuits will accumulate noise, which will impact the quality of the measured estimate in addition to the \textit{shot noise} obtain from the finite sample size.
Several approaches for QAOA-specific error mitigation techniques exist that can markedly improve the estimate's quality.
One prominent approach is to leverage symmetries within the classical objective function, which corresponds to joint degeneracies of the objective Hamiltonian's eigenspaces.
Symmetrizing the trial states $\ket{\vec{\beta}, \vec{\gamma}}$ with respect to these degeneracies can increase the estimate's accuracy~\cite{Shaydulin2021ErrorMitigationForDeepQuantumOptimizationCircuitsByLeveragingProblemSymmetries,Kakkar2022CharacterizingErrorMitigationBySymmetryVerificationInQAOA}.
A second approach is centered around post-selecting directly on the COP's constraints in order to project back to the feasible subspace~\cite{Botelho2022ErrorMitigationForVariationalQuantumAlgorithmsThroughMidCircuitMeasurements}.
Furthermore, and more generally, redundancy encoding of classical variables also enables error mitigation schemes with applications also in combinatorial optimization~\cite{Weidinger2023ErrorMitigationForQuantumApproximateOptimization}.

\subsubsection{Phase separator}

The unitary phase separator is supposed to render the classical objective function's behavior but is technically merely required to be diagonal in the computational basis.
We want to be more precise and suggest the following definition.
\begin{Definition}\label{definition:PhaseSeparator}
    A Hamiltonian $H$ is called a \textit{phase separator Hamiltonian} if it fulfills the following two conditions:
    \begin{itemize}
        \item[(i)] $H$ is diagonal in the computational basis.
        \item[(ii)] The eigenspace of $H\vert_{\mathcal{S}}$ corresponding to its smallest eigenvalue is $\mathcal{S}_{\min}$.
    \end{itemize}
    Then
    \begin{align}
        \phaseu(H,\gamma) \coloneqq e^{-i \gamma H}
    \end{align}
    is the corresponding (\textit{parameteerized}) \textit{phase separator}.
\end{Definition}
The canonical choice for a phase separator Hamiltonian is the objective Hamiltonian $C$.
However, there might be decent approximations of $C$ that are easier to implement and still preserve the optimal solution space, e.g., in threshold-based QAOA~\cite{Golden2021ThresholdBasedQuantumOptimization}.
Since the phase separator is itself diagonal in the computational basis it trivially leaves the solution space $\mathcal{S}$ invariant.

\subsubsection{Mixer}

The unitary mixer is supposed to preserve and explore the solution space $\mathcal{S}$.
Preservation of $\mathcal{S}$ simply means that $\mixu(\beta)(\mathcal{S}) \subseteq \mathcal{S}$ should hold for all $\beta \in \R$.
The exploring condition is defined as follows:
for all $\bm{z}, \bm{z}' \in \mathcal{S}$, there should exist a power $r \in \N$ and a parameter value $\beta \in \R$ so that $\braket{\bm{z} | \mixu^{r}(\beta) | \bm{z}'} \neq 0$.\\
Our aim for a refined definition is now to mimic the two properties of the original QAOA mixer Hamiltonian
\vspace{-6pt}
\begin{align}\label{equation:StandardMixer}
    B = \sum_{n = 1}^{N} \sigma_{x}^{(n)}
\end{align}
for unconstrained problems, but tailored to the constrained case.
$B$, considered as a matrix in the computational basis, is component-wise non-negative and irreducible.
Irreducibility means that the matrix $B$ does not leave any non-trivial \textit{coordinate subspace} invariant.
That is, the only two subspaces of $\hil$ that are a linear span of computational basis states and are left invariant under $B$ are $\{0\}$ and $\hil$.
For brevity, we will address every linear span of computational basis states as a \textit{coordinate subspace}.
Our crucial observation is that the concept of irreducibility is indeed a fundamental mixing property that should be preserved in the constrained case~\cite{Binkowski2024ElementaryProofOfQAOAConvergence}.

In order to establish also `sequential' mixers~\cite{Hadfield2019FromTheQuantumApproximateOptimizationAlgorithmToAQuantumAlternatingOperatorAnsatz}, we utilize the following result, which can be proved by considering each Hamiltonian as the adjacency matrix of a graph.
More precisely, in a first step, one can interpret the matrix representation of each Hamiltonian as the adjacency matrix of a weighted directed graph.
However, since we are not interested in actual weights and all matrices are hermitian, it suffices to consider simply a graph.
Furthermore, since all components are non-negative, no cancellation happens when summing up all the matrices~\cite{Binkowski2022ConstraintGraphModelAnalysisOfTheQuantumAlternatingOperatorAnsatz}.

\begin{Proposition}\label{proposition:MixingProperty}
    Let $\{H_{i}\}_{i \in I} \subset \lo(\hil)$, $0 < \lvert I\rvert < \infty$, be a family of Hamiltonians, component-wise non-negative in the computational basis, such that $H_{i}(\mathcal{S}) \subseteq \mathcal{S}$ holds for all $i \in I$.
    Then the following two statements are equivalent:
    \vspace{-6pt}
    \begin{itemize}
        \item[(i)] Any coordinate subspace $X \subseteq \mathcal{S}$ that is left invariant under every $H_{i}\vert_{\mathcal{S}}$, is trivial.
        \item[(ii)] $\big(\sum_{i \in I} H_{i}\big)\vert_{\mathcal{S}} \in \lo(\mathcal{S})$ is irreducible in the computational basis.
        \vspace{-6pt}
    \end{itemize}
\end{Proposition}

\begin{Definition}\label{definition:Mixer}
    A family of Hamiltonians $\mathsf{H} = \{H_{i}\}_{i \in I} \subset \lo(\hil)$ fulfilling the conditions in \autoref{proposition:MixingProperty} is called a \textit{mixing family}.
    The corresponding (\textit{parameterized}) \textit{simultaneous mixer} is defined as
    \vspace{-6pt}
    \begin{align}
        \simmixu(\mathsf{H}, \beta) \coloneqq e^{- i \beta \sum_{i \in I} H_{i}}.
    \end{align}
    Specifying a permutation $\sigma \in S(I)$, the corresponding (\textit{parameterized}) \textit{sequential mixer} is defined as
    \begin{align}
        \seqmixu(\mathsf{H}, \beta) \coloneqq \prod_{i \in I} e^{- i \beta H_{\sigma(i)}}.
    \end{align}
\end{Definition}

We present here a general method for obtaining suitable mixers from the feasibility-preserving subgroup $F$.
Following the permutation representation, we identify each $g \in F$ via its action on $\bits$ with a linear operator $\rho(g) \in \lo(\hil)$, and via its restricted action on $S(\COP)$ with a linear operator $\rho_{\mathcal{S}}(g) \in \lo(\mathcal{S})$.
Both representations yield permutation matrices in the computational basis.
In the same way we have identified $S(\COP)$ with $\mathcal{S}$, we may also identify every subset of solutions with coordinate subspaces of $\mathcal{S}$.
Then it readily follows that $F$ acts transitively on $S(\COP)$ if the only coordinate subspaces of $\mathcal{S}$ that are left invariant by every $\rho_{\mathcal{S}}(g)$, $g \in F$, are $\{0\}$ and $\mathcal{S}$.
Starting from the unitary operators $\rho(F) = \{W_{g}\}_{g \in F}$, we construct a family of Hamiltonians by taking suitable matrix logarithms $\{i L(W_{g})\}_{g \in F}$.
Since we only have finitely many unitary matrices $W_{g}$, we can always find a common branch $L$ of the complex logarithm for the union of all their eigenvalues.
A direct calculation yields that $A \subseteq \hil$ is a $W$-invariant subspace if $A$ is an $L(W)$-invariant subspace.
Thus, the constructed family $\{i L(W_{g})\}_{g \in F}$ admits the just introduced mixing property whenever $F$ acts transitively on $S(\COP)$ and can therefore be used to build simultaneous and \mbox{sequential mixers.}

As we have shown before, the action of $F$ on $S(\text{OSSP})$ is indeed transitive.
In addition, $F$ consists of bit (value) permutations.
Thus, the operator analogs $\rho(F)$ respect the tensor product structure of the qubit space $\hil$, namely
\vspace{-6pt}
\begin{align}
    \rho(g) \bigotimes_{n = 1}^{N} \ket{\psi_{n}} = \bigotimes_{n = 1}^{N} \ket{\psi_{g^{-1}(n)}}.
\end{align}
Therefore, our formalism yields particularly suitable mixers for the OSSP.
It is, however, definitely not restricted to scheduling-type problem as it solely hinges on the identification and transitive action of a feasibility-preserving group.
Other potential applications are perfect matching with flipping edges along even-length alternating cycles, XOR-SAT with affine transformations, and conservation-constrained network flow with flow in- and decrements along directed cycles.
In general, the notion of a feasibility-preserving group seems to favor problem classes with equality constraints.
Inequality constraints, on the other hand, seem to not induce sufficient symmetries to allow for a group-theoretical treatment.
Take, for example, the knapsack problem where the only constraint is that an inner product of the binary variable vector with some non-negative weight vector should not exceed some positive capacity.
There is no apparent symmetry and hence no apparent invertible operation, which always maps feasible assignments to feasible assignments.
Problems of these types are therefore not suited to profit from our approach.

\subsection{\label{subsection:AQuantumGroupOptimizationAlgorithm}A quantum group optimization algorithm}

After embedding our group-theoretic approach into the established QAOA framework, we now derive an alternative VQA design specifically tailored to busy OSSP instances~\cite{Kossmann2023AQuantumAlgorithmWithGroupTheory}.
Unlike the QAOA, it does not require a phase separator and, most importantly, comes with a natural upper bound for the number of parameters necessary to reach every feasible solution (especially the optimum) with certainty.

According to the previous section, we can describe the entire solution set $S$ of this problem with a bit-to-qubit mapping.
Namely, the set of all the solutions is identified with the symmetric group $S_{J}$.
Fixing one solution $s \in S$, we can consider the problem of optimizing $f$ equivalently as optimizing
\vspace{-6pt}
\begin{align*}
    \tilde{f}: S_{J} \to \R; \quad g \mapsto f({}^g s).
\end{align*}
Thus busy OSSP instances (this also includes TSP) can be cast to optimization problems over symmetric groups.
The underlying group structure can be exploited in the following way:
Consider an arbitrary $\sigma \in S_{J}$.
There is a well-known representation as a product of at most $\frac{J (J - 1)}{2}$ transpositions~\cite{Humphreys1992ReflectionGroupsAndCoxeterGroups}.
We can further write every transposition as a product of some of the $J - 1$ specific transpositions $\tau_{1} = (1, 2)$, $\ldots$, $\tau_{J - 1} = (J - 1, J)$ in $S_{J}$.
In summary, we find for $\sigma \in S_J$ a binary vector $r^{(\sigma)} \in \{0,1\}^{\frac{J(J-1)^2}{2}}$ such that
\vspace{-6pt}
\begin{align}
    \sigma = \prod_{k=1}^{\frac{J(J-1)}{2}} (\tau_1^{r^{(\sigma)}_{k_1}}\cdots \tau_{J-1}^{r^{(\sigma)}_{k_{J-1}}}).
\end{align}
Recall that the elements of $S_{J}$ simultaneously act on multiple vertices, namely those lying in different position blocks, but having the same job coordinate.
The representation of a transposition $\tau_{i}$ as elements in $\lo(\mathcal{H})$ thus yields a product of disjoint SWAP gates, with each SWAP gate corresponding to one position block:
\vspace{-6pt}
\begin{align}
    \tau_{i}\, \widehat{=}\, \prod_{j = 1}^{J} \text{SWAP}_{(i, i + 1)}^{(j)} \eqqcolon B_{i}.
\end{align}
Here, $\text{SWAP}_{(i, i')}^{(j)}$ interchanges the $i$th and the $i'$th qubit in the $j$th position block.
Since all SWAP gates are also hermitian, the operator $B_{i}$ is again hermitian.
Introducing $J (J - 1)^{2} / 2$ parameters and exponentiating each transposition operator $B_{i}$ then yields the following implementation of $\sigma$ as a parameterized \mbox{quantum circuit}:
\vspace{-6pt}
\begin{align}\label{equation:circuit}
    U(\vec{\beta}\,) = \prod_{k = 1}^{\frac{J (J - 1)}{2}} e^{-i \beta_{k_{1}} B_{1}} \cdots e^{-i \beta_{k_{J - 1}} B_{J - 1}}. \vspace{-6pt}
\end{align}
Due to the fact that $e^{-i \beta \text{SWAP}}$ is equal to SWAP with a global phase $i$ for $\beta = -\pi / 2$ and equal to the identity $\one$ for $\beta = 0$, we can precisely implement the action of each permutation $\sigma \in S_{J}$ with the right choice for the $J (J - 1)^{2} / 2$ parameters values.
Therefore, we can directly transfer \autoref{theorem:transitivity} to the gate implementation of permutations in $S_{J}$ to conclude the following theorem:
\begin{Theorem}\label{theorem:PerformanceGuarantee}
    {\addfontfeature{LetterSpace=-4.0}Let $\OSSP(M, T, J)$ be a busy OSSP instance, i.e.,\ $M T = J$.
        Starting from a feasible computational basis state $\ket{\iota} \in \mathcal{S}$, there exist values $\vec{\beta} \in [0, \frac{\pi}{2}]^{\frac{J (J - 1)^{2}}{2}}$ so that applying $U(\vec{\beta})$, as defined in \textbf{(\ref{equation:circuit})}, to $\ket{\iota}$ yields an optimal solution to $\OSSP(M, T, J)$.
        Therefore, sampling from the resultant state returns the optimum with certainty.}
    \vspace{-6pt}
\end{Theorem}

\autoref{theorem:PerformanceGuarantee} constitutes a strong theoretical performance guarantee and has to be contrasted with substantially weaker guarantees for the generic QAOA.
Here, only in the limit $p \to \infty$ one is guaranteed to reach an optimal solution, but also subject to selecting (in this case an entire sequence of) suitable parameter values~\cite{Binkowski2024ElementaryProofOfQAOAConvergence}.
However, the practical implications of \autoref{theorem:PerformanceGuarantee} are ultimately limited by our ability to indeed find optimal parameter values and by quantum hardware limitations such as limited numbers of available qubits, missing qubit connectivity, decoherence, and gate infidelities.
Nevertheless, such theoretical guarantees, even if not entirely transferable to a realistic setup, provide strong guidance for the formulation of practically useful heuristics.
In this particular case, the above result inspires the following VQA:
\vspace{-6pt}
\begin{itemize}
    \item[1.] Start in a solution state $\ket{\iota} \in \mathcal{S}$.
    \item[2.] Apply the parameterized quantum circuit \textbf{(\ref{equation:circuit})}.
    \item[3.] Variationally optimize $\vec{\beta} \in [0, \frac{\pi}{2}]^{\frac{J (J - 1)^{2}}{2}}$ using a classical optimization rule.
    \item[4.] Measure the final outcome state in the computational basis.
\end{itemize}

{\addfontfeature{LetterSpace=-3.5}Furthermore, the parameter space is always the same and especially compact, regardless of the objective function $f$.
    This has to be seen in contrast to general QAOA-mixers for which no comparable restriction of the necessary parameter space is possible.
    Thus, in our case, sampling methods for parameter adaptation like the sampled gradient descent (compare Section 3.7) are particularly powerful since they eventually grant access to the entire compact parameter space.
    However, we highlight that global parameter optimization quickly becomes infeasible for larger values of $J$ due to the effect of shot and hardware noise~\cite{PellowJarman2024TheEffectOfClassicalOptimizersAndAnsatzDepthOnQAOAPerformanceInNoisyDevices}, and because of the general hardness of multi-dimensional, non-convex optimization~\cite{Bittel2021TrainingVariationalQuantumAlgorithmsIsNPHard}.
    More precisely, the parameter optimization landscape typically suffers from exponentially flat regions---a phenomenon called \textit{barren plateaus}~\cite{Larocca2025BarrenPlateausInVariationalQuantumComputing}---and from a large number of local minima~\cite{Kossmann2025DeepCircuitQAOA}.
    On the bright side, there are already several frameworks for mitigating the effect of barren plateaus~\cite{Patti2021EntanglementDevisedBarrenPlateauMitigation,Sack2022AvoidingBarrenPlateausUsingClassicalShadows}, enhancing parameter optimization strategies~\cite{Egger2021WarmStartingQuantumOptimization,Lee2024IterativeLayerwiseTrainingForTheQuantumApproximateOptimizationAlgorithm}, and formulating the parameter optimization as an easier problem~\cite{Binkowski2025FromBarrenPlateausThroughFertileValleysConicExtensionsOfParameterisedQuantumCircuits,Binkowski2024OneForAllUniversalQuantumConicProgrammingFrameworkForHardConstrainedCombinatorialOptimizationProblems} that are applicable to the general VQAs, including our algorithm.
    Moreover, note that the aforementioned error mitigation techniques are also applicable to our algorithmic scheme:
    redundancy encoding of binary variables is a generally valid strategy to introduce error mitigation techniques.
    The notion of feasibility is easily updated for redundantly encoded variables, merely resulting in additional SWAP gates within $B_{i}$ circuits.
    While our linear OSSP objective function \textbf{(\ref{equation:ObjectiveFunction})} will typically not have symmetries to exploit for error mitigation purposes, the notion of OSSP feasibility definitely falls within the range of feasibility-based error mitigation protocols considered in~\cite{Botelho2022ErrorMitigationForVariationalQuantumAlgorithmsThroughMidCircuitMeasurements}.}

The identification of the set of feasible solutions with $S_{J}$ further suggests a dynamic programming~\cite{Cormen2022IntroductionToAlgorithms2022} ansatz:
Iteratively optimize over subgroup series of the form
\vspace{-6pt}
\begin{align}
    1 = \langle \tau_{i} \defcolon i \in I_{0} \rangle \leq \ldots \leq \langle \tau_{i} \defcolon i \in I_{n} \rangle = S_{J},
\end{align}
with ascending index sets $I_{0} \subset \ldots \subset I_{n}$, by restricting to the set of corresponding mixers $B_{i}$ with $i \in I_{k}$, $k \in [n]$.

{\addfontfeature{LetterSpace=-2.0}Lastly, let us discuss a concrete decomposition scheme of the compound gate $U(\vec{\beta})$ into logical components.
    It consists of a product of $J (J - 1)^{2} / 2$ exponentials of the Hamiltonians $B_{i}$ such that it suffices to focus on the decomposition of a single exponential.
    The entire implementation of $U(\vec{\beta})$ is then given by sequential execution of each individually decomposed exponential.
    A single $B_{i}$ consists of $J$ SWAP gates on mutually disjoint qubit pairs.
    SWAP gates are involuntary, i.e.,\ $\text{SWAP}^{2} = \one$, which allows us to exactly implement their exponential with the aid of a single ancilla qubits via the circuit depicted in \textbf{\autoref{figure:ExponentialDecomposition}};
    a thorough proof can be found in~\cite[Theorem 5.25]{Schwiering2024QuantumOptimizationAlgorithmsForTheTravelingSalesmanProblem}.
    All components of this circuit except for the controlled $B_{i}$-operations are elementary single-qubit gates, hence adding only a constant overhead in both gate count and circuit depth.
    The entire construction essentially boils down to implementing a singly controlled product of SWAP gates, i.e.,\ a product of Fredkin gates.
    Even though all SWAP gates contained in $B_{i}$ act on mutually disjoint pairs of qubits and are thus parallelizable (assuming the quantum hardware supports parallel execution of gates on disjoint qubits), this does not hold anymore for the controlled version, as all SWAP gates are controlled on the same qubit.
    Furthermore, Fredkin gates do not usually belong to a quantum computer's native gate set.
    They can be logically decomposed into a Toffoli gate and two enclosing CNOT gates.
    The Toffoli gate, in turn, has an optimal decomposition into nine single-qubit gates and six CNOT gates~\cite{Shende2009OnTheCNOTCostOfTOFFOLIGates}, introducing only a constant implementation overhead.
    In summary, decomposing the entire circuit $U(\beta)$ into single-qubit and CNOT gates requires $\mathcal{O}(J^{4})$ elementary gates.}

\begin{figure*}[!ht]
    \centering
    \begin{quantikz}[row sep=0.4cm, column sep=0.3cm, align equals at=1.5]
        \lstick{$\ket{\psi}$} & \gate{e^{-i \beta B_{i}}} & \qw \\
        \lstick{$\ket{0}$} & \hphantomgate{R_{X}(2 \beta)} & \qw
    \end{quantikz}
    =
    \begin{quantikz}[row sep=0.4cm, column sep=0.3cm, align equals at=1.5]
        \lstick{$\ket{\psi}$} & \hphantomgate{e^{-i \beta B_{i}}} & \gate{B_{i}} & \qw & \gate{B_{i}} & \qw & \qw \\
        \lstick{$\ket{0}$} & \gate{H} & \ctrl{-1} & \gate{R_{X}(2 \beta)} & \ctrl{-1} & \gate{H} & \qw
    \end{quantikz}
    \caption{\label{figure:ExponentialDecomposition}
        Quantum circuit implementing the parameterized exponential of $B_{i}$.
        This construction requires a $\ket{0}$-initialized temporary working qubit, that is, the additional qubit returns to the $\ket{0}$ with certainty.
        The parameter $\beta$ is introduced via a parametrized $X$-rotation on the ancilla qubit.
        The entire construction only depends on $B_{i}$ being involuntary, i.e.,\ $B_{i}^{2} = \one$.
    }
\end{figure*}

\subsection{\label{subsection:NumericalResultsExactSimulation}Numerical results: exact simulation}

First, we demonstrate our just introduced VQA in a noiseless (neither shot- nor hardware noise) classical simulation on an OSSP(2,2,4) instance.
\textbf{\autoref{figure:OSSP224Graph}} displays this instance's constraint graph.
The group of job permutations is thus given by $S_{J} = S_{4}$.
After assigning a bit to each of the 16 elements in $[2] \times [2] \times [4]$ via $(m, t, j) \mapsto 4 (m - 1) + 2 (t - 1) + j$, we explicitly generate $S_{J}$ with $\tau_{1}$, $\tau_{2}$, and $\tau_{3}$:
\begin{align}
    \begin{split}
        S_{J} = \langle (1,2)(5,6)(9,10)(13,14), (2,3)(6,7)(10,11)(14,15),\\
        (3,4)(7,8)(11,12)(15,16)\rangle.
    \end{split}
\end{align}
Consequently, we have to implement three distinct mixer Hamiltonians $B_{1}$, $B_{2}$, $B_{3}$.
As a concrete example, consider the permutation $(1,2)(5,6)(9,10)(13,14)$, which corresponds to the Hamiltonian
\begin{align}
    B_{1}\! \!= \! \!\sum_{p = 1}^{4}\! \SWAP_{(1, 2)}^{(p)}\! \!=\! \!\SWAP_{(1,2)}\! \SWAP_{(5,6)}\! \SWAP_{(9,10)}\! \SWAP_{(13,14)}.
\end{align}
As a concrete feasible initial state, we choose $\ket{\bm{z}_0} = \ket{1000010000100001}$.

Since $J = 4$, our variational circuit \textbf{(\ref{equation:circuit})} contains 18 parameters to be optimized.
At this scale, global black-box optimization techniques such as differential evolution~\cite{Storn1997DifferentialEvolutionASimpleAndEfficientHeuristicForGlobalOptimizationOverContinuousSpaces} already becomes unpractical and we have to resort to local optimization strategies.
Concretely, we utilize the L-BFGS-B algorithm~\cite{Byrd1995ALimitedMemoryAlgorithmForBoundConstrainedOptimization} as implemented in scipy.
However, in order to overcome the local optimizer's tendency to get stuck in a local minimum and to avoid the risk of initializing the parameters within a barren plateau, we enhance the parameter optimization in the following two ways:
First, we iteratively optimize first $2 q$ parameters, $1 \leq q \leq 9$ with the first $2 (q - 1)$ parameters being warm-started from the prior iteration.
For $q = 9$ this eventually covers all $18$ parameters, but with quantitative warm starts for the first $16$ parameters, rather than random initialization.
At each iteration step, we include two additional parameters into the scope of optimization.
Our second enhancement consists of initializing the additional parameters with several value pairs, stemming from a shared grid.
We conduct the subsequent optimization for each chosen pair of initial parameter values and only keep the best performing one.

We further compare the performance of our VQA with the standard QAOA and softcoded constraints.
Recall, that softcoding the constraints means to alter the objective function $f$ in order to penalize infeasible solutions.
We adapt the standard penalization strategy for Hamiltonian cycles~\cite{Lucas2014IsingFormulationsOfManyNPProblems} with penalty function
\begin{align}
    g(\bm{z}) \coloneqq \sum_{m = 1}^{M} \sum_{t = 1}^{T} \bigg(1 - \sum_{j = 1}^{J} z_{m t j}\bigg)^{2}\! +\! \sum_{j = 1}^{J} \bigg(1 - \sum_{m = 1}^{M} \sum_{t = 1}^{T} z_{m t j}\bigg)^{2}.
\end{align}
The adapted objective function then reads $f + \alpha g$, where $\alpha$ is chosen so that no infeasible bit string receives a value lower or equal to the minimum of $f$ over the feasible solutions.
In reality, one can only guess or give non-tight upper bounds for $\alpha$.
For our toy example, however, we compute the optimal, i.e.,\ lowest possible value for $\alpha$ by looping through all $2^{16}$ bit strings and determining the smallest value for $\alpha$, which lifts all infeasible bit strings above the optimal feasible one.
The QAOA circuit consists of an alternating application of the exponentiated standard mixer \textbf{(\ref{equation:StandardMixer})} and the usual phase separator, and is applied to $\ket{+}$ as initial state.
For better comparability, we implement the QAOA with depth $9$ and optimize its $18$ variational parameters with the exact same strategy as described above for our feasibility-preserving VQA.

\begin{figure*}[!ht]
    \centering
    \includegraphics[width=1\linewidth]{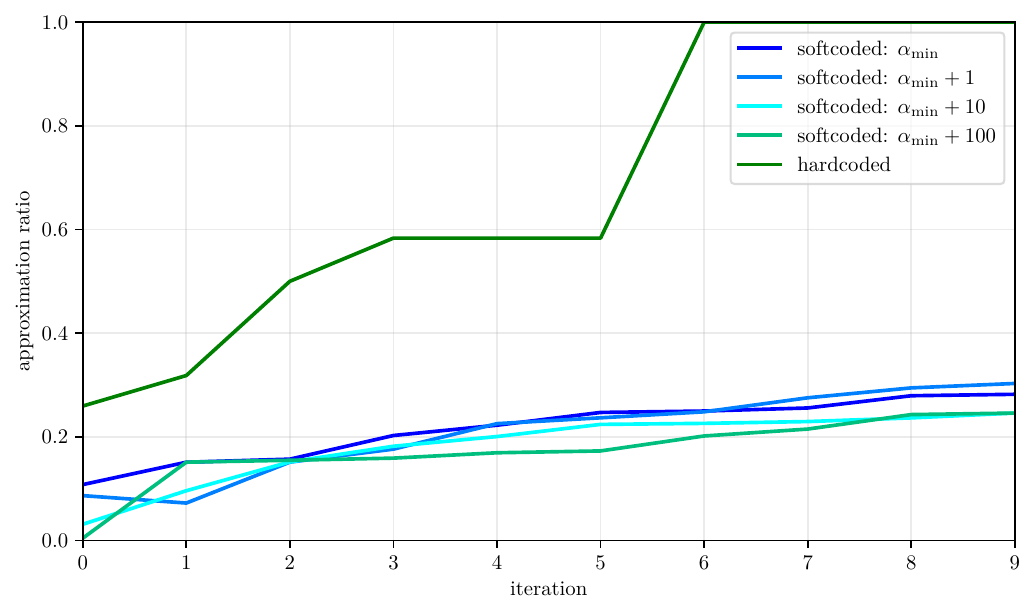}
    \caption{\label{figure:NoiselessExperiment}
        Numerical experiments for the OSSP(2,2,4) instance.
        The plots depicts the approximation ratios (theoretical minimum divided by expectation value) after each iteration of parameter optimization for the standard QAOA with four different penalties as well as our VQA with hardcoded constraints.
        We observe that our VQA is able to reach the optimum (approximation ratio of $100\%$) already after six iterations, which corresponds to 12 optimized parameters.
        In contrast, the standard QAOA is not able to produce approximation ratios above $30\%$, even when having access to all $18$ variational parameters.
    }
\end{figure*}

The results of our noiseless are depicted in \textbf{\autoref{figure:NoiselessExperiment}}.
Note that the approximation ratios are always with respect to the chosen $\alpha$ value.
Accordingly, the initial state $\ket{+}$ (the state obtained after 0 iterations) exhibits different approximation values.
Since our VQA stays within the feasible subspace, different values for $\alpha$ do not affect its achieved approximation ratios.
We clearly observe that the standard approach with softcoded constraints fails to reliably deliver high-quality, feasible solutions.
For the tested range of $\alpha$-values, it merely achieves approximation ratios up to $30\%$.
In comparison, starting from a feasible computational basis state and hardcoding the constraints within our VQA yields better ratios across all iterations.
In fact, for this specific example we observe that optimizing the first $12$ parameters already suffices to reach the global minimum.
For larger values of $J$, jointly optimizing all $J (J - 1)^{2} / 2$ parameters in the last step will eventually become unpractical.
However, optimizing smaller subsets of parameters in our VQA still holds the potential to significantly improve the initial solution.
The drastic performance differences between our VQA with hardcoded constraints and the standard QAOA with softcoded constraints have two simple explanations, which also generalize to larger instance sizes:
First, our design comes with a guaranteed upper bound of parameters needed to achieve all possible solutions with certainty, including the optimum (see \autoref{theorem:PerformanceGuarantee}), while for the generic QAOA there are no such guarantees.
Second, the feasible subspace is of much lower dimension than the entire Hilbert space.
For $J = 24$, there are $4! = 24$ feasible solutions, but $2^{4^{2}} = 65536$ different bit strings in total, and the ratio of feasible solutions to arbitrary bit strings quickly decreases further for larger numbers of $J$.
Accordingly, the standard QAOA initial state $\ket{+}$ only has very small overlap with the feasible states and the guidance via softcoded constraints is not sufficient to concentrate enough amplitude within the feasible subspace.

\subsection{\label{subsection:NumericalResultsRealHardwareImplementation}Numerical results: real hardware implementation}

As another proof of principle, we consider an OSSP(1,3,3) instance, i.e.,\ a TSP instance with three cities, and demonstrate our VQA on a real (noisy) quantum hardware: the IBM Q System One.
For the same problem instance, we also conduct a noisy simulation of the VQA on a classical computer.

The group of job permutations is given by $S_{J} = S_{3}$.
Since we only have one machine, we simply drop the machine coordinate and consider the enumeration $(t, j) \mapsto 3 (t - 1) + j$.
Thus, $S_{J}$ is generated by two elements $\tau_{1} = (1,2)(4,5)(7,8)$ and $\tau_{2} = (2,3)(5,6)(8,9)$.
We further consider the objective function
\begin{align}
    f:\{0,1\}^{9} \rightarrow \R,\quad \bm{z} \mapsto \sum_{t, j} \omega_{t j} z_{t j}\ \\
    \text{ with weight matrix } (\omega)_{t j} = \left(
    \begin{array}{rrr}
        3 & 2 & 2 \\
        2 & 2 & 3 \\
        1 & 2 & 2 \\
    \end{array}
    \right).
\end{align}

\begin{figure*}[!hb]
    \hspace{30pt}
    \begin{minipage}[t]{0.427\linewidth}
        \centering
        \vspace{2cm}
        \begin{tabular}{|c|c|}
            \hline
            \text{bit string} & \text{value} \\
            \hline
            001010100 & 5 \\
            001100010 & 6 \\
            010001100 & 6 \\
            010100001 & 7 \\
            100001010 & 8 \\
            100010001 & 8 \\
            \hline
        \end{tabular}
    \end{minipage}\hfill
    \begin{minipage}[t]{0.573\linewidth}
        \vspace{0pt}
        \begin{tikzpicture}[scale=0.1]
            \tikzset{
                bitnode/.style={
                    rotate=80
                },
                simnode/.style={
                    scale=0.1,
                    anchor=south,
                    fill=green,
                    minimum width=2.2cm,
                    draw=black,
                    very thin,
                },
                ibmnode/.style={
                    scale=0.1,
                    anchor=south,
                    fill=blue,
                    minimum width=2.2cm,
                    draw=black,
                    very thin,
                }
            }

            \node (a) at (0,71) {\textbf{1.}};

            \node[bitnode] (b1) at (0,-1) {010001001};
            \node[bitnode] (b2) at (5,-1) {010001010};
            \node[bitnode] (b3) at (10,-1) {010001100};
            \node[bitnode] (b4) at (15,-1) {010010001};
            \node[bitnode] (b5) at (20,-1) {010100001};
            \node[bitnode] (b6) at (25,-1) {100001001};
            \node[bitnode] (b7) at (30,-1) {100001010};
            \node[bitnode] (b8) at (35,-1) {100001100};
            \node[bitnode] (b9) at (40,-1) {100010001};
            \node[bitnode] (b10) at (45,-1) {100010100};
            \node[bitnode] (b11) at (50,-1) {100100001};
            \node[bitnode] (b12) at (55,-1) {rest};

            \draw (-4.5,9) --++(62.5,0);

            \draw (0,8.5) --++(0,1);
            \draw (5,8.5) --++(0,1);
            \draw (10,8.5) --++(0,1);
            \draw (15,8.5) --++(0,1);
            \draw (20,8.5) --++(0,1);
            \draw (25,8.5) --++(0,1);
            \draw (30,8.5) --++(0,1);
            \draw (35,8.5) --++(0,1);
            \draw (40,8.5) --++(0,1);
            \draw (45,8.5) --++(0,1);
            \draw (50,8.5) --++(0,1);
            \draw (55,8.5) --++(0,1);

            \draw (-4,9) --++(0,65);
            \node (y1) at (-6,9) {0};
            \draw[gray,dashed] (-4,24) --++(62,0);
            \draw (-4.5,24) --++(1,0);
            \node (y2) at (-8,24) {150};
            \draw[gray,dashed] (-4,39) --++(62,0);
            \draw (-4.5,39) --++(1,0);
            \node (y3) at (-8,39) {300};
            \draw[gray,dashed] (-4,54) --++(62,0);
            \draw (-4.5,54) --++(1,0);
            \node (y4) at (-8,54) {450};
            \draw[gray,dashed] (-4,69) --++(62,0);
            \draw (-4.5,69) --++(1,0);
            \node (y5) at (-8,69) {600};

            \node[rotate=90,anchor=south] (ylabel) at (-11,39) {counts};

            \node[simnode,minimum height=125mm] (s1) at (-1.1,9) {};
            \node[ibmnode,minimum height=70mm] (i1) at (1.1,9) {};

            \node[simnode,minimum height=50mm] (s2) at (3.9,9) {};
            \node[ibmnode,minimum height=39mm] (i2) at (6.1,9) {};

            \node[simnode,minimum height=65mm] (s3) at (8.9,9) {};

            \node[simnode,minimum height=89mm] (s4) at (13.9,9) {};
            \node[ibmnode,minimum height=36mm] (i4) at (16.1,9) {};

            \node[simnode,minimum height=85mm] (s5) at (18.9,9) {};
            \node[ibmnode,minimum height=35mm] (i5) at (21.1,9) {};

            \node[simnode,minimum height=121mm] (s6) at (23.9,9) {};
            \node[ibmnode,minimum height=80mm] (i6) at (26.1,9) {};

            \node[simnode,minimum height=43mm] (s7) at (28.9,9) {};
            \node[ibmnode,minimum height=35mm] (i7) at (31.1,9) {};

            \node[simnode,minimum height=52mm] (s8) at (33.9,9) {};
            \node[ibmnode,minimum height=27mm] (i8) at (36.1,9) {};

            \node[simnode,minimum height=66mm] (s9) at (38.9,9) {};
            \node[ibmnode,minimum height=47mm] (i9) at (41.1,9) {};

            \node[ibmnode,minimum height=26mm] (i10) at (46.1,9) {};

            \node[simnode,minimum height=75mm] (s11) at (48.9,9) {};
            \node[ibmnode,minimum height=29mm] (i11) at (51.1,9) {};

            \node[simnode,minimum height=253mm] (s12) at (53.9,9) {};
            \node[ibmnode,minimum height=600mm] (i12) at (56.1,9) {};

            \node[fill=green,minimum width=6mm,draw=black,very thin,anchor=west] (simlegend) at (-1,63.5) {};
            \node[anchor=west] (simlegendtext) at (6,63.5) {simulator};
            \node[fill=blue,minimum width=6mm,draw=black,very thin,anchor=west] (ibmlegend) at (-1,59.5) {};
            \node[anchor=west] (ibmlegendtext) at (6,59.5) {IBM Q System One};
            \draw[draw=black,rounded corners] (-2.5,57) rectangle ++ (47,9);
        \end{tikzpicture}
    \end{minipage}\hfill
    \begin{minipage}[t]{0.535\linewidth}
        \hspace{30pt}
        \begin{tikzpicture}[scale=0.1]
            \tikzset{
                bitnode/.style={
                    rotate=80
                },
                simnode/.style={
                    scale=0.1,
                    anchor=south,
                    fill=green,
                    minimum width=2.2cm,
                    draw=black,
                    very thin,
                },
                ibmnode/.style={
                    scale=0.1,
                    anchor=south,
                    fill=blue,
                    minimum width=2.2cm,
                    draw=black,
                    very thin,
                }
            }

            \node (a) at (0,71) {\textbf{2.}};

            \node[bitnode] (b1) at (0,-1) {010001001};
            \node[bitnode] (b2) at (5,-1) {010001010};
            \node[bitnode] (b3) at (10,-1) {010001100};
            \node[bitnode] (b4) at (15,-1) {010010001};
            \node[bitnode] (b5) at (20,-1) {010010010};
            \node[bitnode] (b6) at (25,-1) {010100001};
            \node[bitnode] (b7) at (30,-1) {010100010};
            \node[bitnode] (b8) at (35,-1) {010100100};
            \node[bitnode] (b9) at (40,-1) {100001001};
            \node[bitnode] (b10) at (45,-1) {100001100};
            \node[bitnode] (b11) at (50,-1) {100100001};
            \node[bitnode] (b12) at (55,-1) {rest};

            \draw (-4.5,9) --++(62.5,0);

            \draw (0,8.5) --++(0,1);
            \draw (5,8.5) --++(0,1);
            \draw (10,8.5) --++(0,1);
            \draw (15,8.5) --++(0,1);
            \draw (20,8.5) --++(0,1);
            \draw (25,8.5) --++(0,1);
            \draw (30,8.5) --++(0,1);
            \draw (35,8.5) --++(0,1);
            \draw (40,8.5) --++(0,1);
            \draw (45,8.5) --++(0,1);
            \draw (50,8.5) --++(0,1);
            \draw (55,8.5) --++(0,1);

            \draw (-4,9) --++(0,65);
            \node (y1) at (-6,9) {0};
            \draw[gray,dashed] (-4,24) --++(62,0);
            \draw (-4.5,24) --++(1,0);
            \node (y2) at (-8,24) {150};
            \draw[gray,dashed] (-4,39) --++(62,0);
            \draw (-4.5,39) --++(1,0);
            \node (y3) at (-8,39) {300};
            \draw[gray,dashed] (-4,54) --++(62,0);
            \draw (-4.5,54) --++(1,0);
            \node (y4) at (-8,54) {450};
            \draw[gray,dashed] (-4,69) --++(62,0);
            \draw (-4.5,69) --++(1,0);
            \node (y5) at (-8,69) {600};

            \node[rotate=90,anchor=south] (ylabel) at (-11,39) {counts};

            \node[simnode,minimum height=171mm] (s1) at (-1.1,9) {};
            \node[ibmnode,minimum height=93mm] (i1) at (1.1,9) {};

            \node[simnode,minimum height=50mm] (s2) at (3.9,9) {};
            \node[ibmnode,minimum height=26mm] (i2) at (6.1,9) {};

            \node[simnode,minimum height=108mm] (s3) at (8.9,9) {};
            \node[ibmnode,minimum height=39mm] (i3) at (11.1,9) {};

            \node[simnode,minimum height=46mm] (s4) at (13.9,9) {};
            \node[ibmnode,minimum height=47mm] (i4) at (16.1,9) {};

            \node[ibmnode,minimum height=17mm] (i5) at (21.1,9) {};

            \node[simnode,minimum height=98mm] (s6) at (23.9,9) {};
            \node[ibmnode,minimum height=64mm] (i6) at (26.1,9) {};

            \node[simnode,minimum height=54mm] (s7) at (28.9,9) {};

            \node[simnode,minimum height=85mm] (s8) at (33.9,9) {};
            \node[ibmnode,minimum height=37mm] (i8) at (36.1,9) {};

            \node[simnode,minimum height=88mm] (s9) at (38.9,9) {};
            \node[ibmnode,minimum height=52mm] (i9) at (41.1,9) {};

            \node[simnode,minimum height=57mm] (s10) at (43.9,9) {};
            \node[ibmnode,minimum height=24mm] (i10) at (46.1,9) {};

            \node[simnode,minimum height=58mm] (s11) at (48.9,9) {};
            \node[ibmnode,minimum height=41mm] (i11) at (51.1,9) {};

            \node[simnode,minimum height=209mm] (s12) at (53.9,9) {};
            \node[ibmnode,minimum height=584mm] (i12) at (56.1,9) {};

            \node[fill=green,minimum width=6mm,draw=black,very thin,anchor=west] (simlegend) at (-1,63.5) {};
            \node[anchor=west] (simlegendtext) at (6,63.5) {simulator};
            \node[fill=blue,minimum width=6mm,draw=black,very thin,anchor=west] (ibmlegend) at (-1,59.5) {};
            \node[anchor=west] (ibmlegendtext) at (6,59.5) {IBM Q System One};
            \draw[draw=black,rounded corners] (-2.5,57) rectangle ++ (47,9);
        \end{tikzpicture}
    \end{minipage}\hfill
    \begin{minipage}[t]{0.5\linewidth}
        \hspace{-0pt}
        \vspace{0pt}
        \begin{tikzpicture}[scale=0.1]
            \tikzset{
                bitnode/.style={
                    rotate=80
                },
                simnode/.style={
                    scale=0.1,
                    anchor=south,
                    fill=green,
                    minimum width=2.2cm,
                    draw=black,
                    very thin,
                },
                ibmnode/.style={
                    scale=0.1,
                    anchor=south,
                    fill=blue,
                    minimum width=2.2cm,
                    draw=black,
                    very thin,
                }
            }

            \node (a) at (0,71) {\textbf{3.}};

            \node[bitnode] (b1) at (0,-1) {010001001};
            \node[bitnode] (b2) at (5,-1) {010001010};
            \node[bitnode] (b3) at (10,-1) {010001100};
            \node[bitnode] (b4) at (15,-1) {010010001};
            \node[bitnode] (b5) at (20,-1) {010100001};
            \node[bitnode] (b6) at (25,-1) {010100010};
            \node[bitnode] (b7) at (30,-1) {010100100};
            \node[bitnode] (b8) at (35,-1) {010101100};
            \node[bitnode] (b9) at (40,-1) {011001100};
            \node[bitnode] (b10) at (45,-1) {100001001};
            \node[bitnode] (b11) at (50,-1) {100001100};
            \node[bitnode] (b12) at (55,-1) {rest};

            \draw (-4.5,9) --++(62.5,0);

            \draw (0,8.5) --++(0,1);
            \draw (5,8.5) --++(0,1);
            \draw (10,8.5) --++(0,1);
            \draw (15,8.5) --++(0,1);
            \draw (20,8.5) --++(0,1);
            \draw (25,8.5) --++(0,1);
            \draw (30,8.5) --++(0,1);
            \draw (35,8.5) --++(0,1);
            \draw (40,8.5) --++(0,1);
            \draw (45,8.5) --++(0,1);
            \draw (50,8.5) --++(0,1);
            \draw (55,8.5) --++(0,1);

            \draw (-4,9) --++(0,65);
            \draw[gray,dashed] (-4,24) --++(62,0);
            \draw (-4.5,24) --++(1,0);
            \draw[gray,dashed] (-4,39) --++(62,0);
            \draw (-4.5,39) --++(1,0);
            \draw[gray,dashed] (-4,54) --++(62,0);
            \draw (-4.5,54) --++(1,0);
            \draw[gray,dashed] (-4,69) --++(62,0);
            \draw (-4.5,69) --++(1,0);

            \node[simnode,minimum height=196mm] (s1) at (-1.1,9) {};
            \node[ibmnode,minimum height=92mm] (i1) at (1.1,9) {};

            \node[simnode,minimum height=60mm] (s2) at (3.9,9) {};
            \node[ibmnode,minimum height=23mm] (i2) at (6.1,9) {};

            \node[simnode,minimum height=230mm] (s3) at (8.9,9) {};
            \node[ibmnode,minimum height=112mm] (i3) at (11.1,9) {};

            \node[simnode,minimum height=33mm] (s4) at (13.9,9) {};
            \node[ibmnode,minimum height=19mm] (i4) at (16.1,9) {};

            \node[simnode,minimum height=101mm] (s5) at (18.9,9) {};
            \node[ibmnode,minimum height=42mm] (i5) at (21.1,9) {};

            \node[simnode,minimum height=41mm] (s6) at (23.9,9) {};

            \node[simnode,minimum height=114mm] (s7) at (28.9,9) {};
            \node[ibmnode,minimum height=36mm] (i7) at (31.1,9) {};

            \node[ibmnode,minimum height=18mm] (i8) at (36.1,9) {};

            \node[ibmnode,minimum height=18mm] (i9) at (41.1,9) {};

            \node[simnode,minimum height=46mm] (s10) at (43.9,9) {};
            \node[ibmnode,minimum height=32mm] (i10) at (46.1,9) {};

            \node[simnode,minimum height=65mm] (s11) at (48.9,9) {};
            \node[ibmnode,minimum height=37mm] (i11) at (51.1,9) {};

            \node[simnode,minimum height=138mm] (s12) at (53.9,9) {};
            \node[ibmnode,minimum height=595mm] (i12) at (56.1,9) {};

            \node[fill=green,minimum width=6mm,draw=black,very thin,anchor=west] (simlegend) at (-1,63.5) {};
            \node[anchor=west] (simlegendtext) at (6,63.5) {simulator};
            \node[fill=blue,minimum width=6mm,draw=black,very thin,anchor=west] (ibmlegend) at (-1,59.5) {};
            \node[anchor=west] (ibmlegendtext) at (6,59.5) {IBM Q System One};
            \draw[draw=black,rounded corners] (-2.5,57) rectangle ++ (47,9);
        \end{tikzpicture}
    \end{minipage}
    \caption{\label{figure:FirstSteps}
        Numerical experiment for the OSSP(1,3,3) instance.
        The table enumerates all feasible solutions.
        The three subplots show the distribution of sampling 1024 times from the quantum computer, resp. the simulated state, after one, two, and three iterations of sampled gradient descent.
        In each plot, we depict the counts for the ten most sampled bit strings in lexicographical order and collect all remaining counts into a ``rest'' state.
        The VQA quickly leaves the initial state $\bm{z}_{0} = 100010001$ already in the \textbf{1.} step
        The initial state is barely sampled already after the \textbf{2.} step, neither on the quantum device nor in the noisy simulation, and reduces further in frequency with the \textbf{3.} step.
        In contrast, the overlap with the local minimum steadily increases from step to step in both cases.\\
        ~\\
        ~\\
        ~\\
    }
\end{figure*}

In addition to the two mixer Hamiltonians $B_{1}$ and $B_{2}$, we also implement the QAOA-phase separator, resulting in the parameterized quantum circuit
{\fontsize{5.5pt}{5.5pt}
    \begin{align}\label{equation:Circuit2}
        U(\vec{\beta},\vec{\gamma}) \!=\! &\underbrace{e^{-i \beta_{1} B_{1}} e^{-i \beta_{2} B_{2}} e^{-i \gamma_{1} C}}_{1} \underbrace{e^{-i \beta_{3} B_{1}} e^{-i \beta_{4} B_{2}} e^{-i \gamma_{2} C}}_{2} \underbrace{e^{-i \beta_{5} B_{1}} e^{-i \beta_{6} B_{2}} e^{-i \gamma_{3} C}}_{3}.
    \end{align}}
However, this circuit is yet too deep to be fully implemented on the quantum device.
Therefore, we restrict our circuit to the first factor of \textbf{(\ref{equation:Circuit2})}.
Choosing the feasible initial state $\ket{\bm{z}_{0}} = \ket{100010001}$, we can predict which feasible states are actually accessible in this setting.
Restricting to one factor in \textbf{(\ref{equation:Circuit2})} classically corresponds to having only access to four group elements: $\id$, $\tau_{1}$, $\tau_{2}$, and $\tau_{1} \tau_{2}$.
Their application to $\bm{z}_{0}$ yields
\begin{align}
    {}^{\id}\bm{z}_{0} = \bm{z}_{0},\ {}^{\tau_{1}}\bm{z}_{0} = 010100001, {}^{\tau_{2}}\bm{z}_{0} = 100001010,\\
    {}^{\tau_{1} \tau_{2}}\bm{z}_{0} = 010001100,
\end{align}
where the colored bit string ${}^{\tau_{1} \tau_{2}}\bm{z}_{0}$ is the optimal accessible feasible solution.
In contrast, the global minimum 001010100 requires at least access to the first two factors as it is readily given by ${}^{\tau_{2} \tau_{1} \tau_{2}}\bm{z}_{0}$.

For the actual parameter adaptation we use sampled gradient descent:
\begin{itemize}
    \item[1.] Construct a ball $\mathbb{B}_{i} \subset [0, \pi / 2]^{3}$ around some parameter values $\vec{\beta}_{i}$.
    \item[2.] Sample new parameter values uniformly from $\mathbb{B}_{i}$.
    \item[3.] Apply the correspondingly parameterized circuit to the initial state and estimate the expectation value of $C$.
    \item[4.] Choose parameter values $\vec{\beta}_{i + 1}$ from the sample minimizing the expectation value and repeat step 1 with $i \mapsto i + 1$.
\end{itemize}
The size of the constructed ball $\mathbb{B}_{i}$ is adapted in each step $i$: The steeper the drop in expectation values, the smaller the radius becomes.
For our numerical execution we choose random initial parameters and a sample size of 40 in each step.
\newpage

\begin{figure*}[!ht]
    \begin{minipage}[t]{0.5\linewidth}

        \centering
        \vspace{0pt}
        \begin{tikzpicture}[scale=0.1]
            \tikzset{
                bitnode/.style={
                    rotate=80
                },
                simnode/.style={
                    scale=0.1,
                    anchor=south,
                    fill=green,
                    minimum width=4.2cm,
                    draw=black,
                    very thin,
                },
                ibmnode/.style={
                    scale=0.1,
                    anchor=south,
                    fill=blue,
                    minimum width=4.2cm,
                    draw=black,
                    very thin,
                }
            }

            \node (a) at (-2,65) {\textbf{4.}};

            \node[bitnode] (b1) at (0,-1) {010001001};
            \node[bitnode] (b2) at (9,-1) {010001010};
            \node[bitnode] (b3) at (18,-1) {010001100};
            \node[bitnode] (b4) at (27,-1) {010100001};
            \node[bitnode] (b5) at (36,-1) {010100100};
            \node[bitnode] (b6) at (45,-1) {011001100};
            \node[bitnode] (b7) at (54,-1) {rest};

            \draw (-6.5,9) --++(66.5,0);

            \draw (0,8.5) --++(0,1);
            \draw (9,8.5) --++(0,1);
            \draw (18,8.5) --++(0,1);
            \draw (27,8.5) --++(0,1);
            \draw (36,8.5) --++(0,1);
            \draw (45,8.5) --++(0,1);
            \draw (54,8.5) --++(0,1);

            \draw (-6,9) --++(0,72);
            \node (y1) at (-8,9) {0};
            \draw[gray,dashed] (-6,24) --++(65,0);
            \draw (-6.5,24) --++(1,0);
            \node (y2) at (-10,24) {150};
            \draw[gray,dashed] (-6,39) --++(65,0);
            \draw (-6.5,39) --++(1,0);
            \node (y3) at (-10,39) {300};
            \draw[gray,dashed] (-6,54) --++(65,0);
            \draw (-6.5,54) --++(1,0);
            \node (y4) at (-10,54) {450};
            \draw[gray,dashed] (-6,69) --++(65,0);
            \draw (-6.5,69) --++(1,0);
            \node (y5) at (-10,69) {600};

            \node[rotate=90,anchor=south] (ylabel) at (-13,39) {counts};

            \node[simnode,minimum height=179mm] (s1) at (-2.1,9) {};
            \node[ibmnode,minimum height=75mm] (i1) at (2.1,9) {};

            \node[simnode,minimum height=46mm] (s2) at (6.9,9) {};

            \node[simnode,minimum height=398mm] (s3) at (15.9,9) {};
            \node[ibmnode,minimum height=140mm] (i3) at (20.1,9) {};

            \node[simnode,minimum height=86mm] (s4) at (24.9,9) {};
            \node[ibmnode,minimum height=37mm] (i4) at (29.1,9) {};

            \node[simnode,minimum height=142mm] (s5) at (33.9,9) {};
            \node[ibmnode,minimum height=50mm] (i5) at (38.1,9) {};

            \node[ibmnode,minimum height=37mm] (i6) at (47.1,9) {};

            \node[simnode,minimum height=173mm] (s7) at (51.9,9) {};
            \node[ibmnode,minimum height=685mm] (i7) at (56.1,9) {};

            \node[fill=green,minimum width=6mm,draw=black,very thin,anchor=west] (simlegend) at (-2,77.5) {};
            \node[anchor=west] (simlegendtext) at (4,77.5) {simulator};
            \node[fill=blue,minimum width=6mm,draw=black,very thin,anchor=west] (ibmlegend) at (-2,73.5) {};
            \node[anchor=west] (ibmlegendtext) at (4,73.5) {IBM Q System One};
            \draw[draw=black,rounded corners] (-4.5,71) rectangle ++ (47,9);
        \end{tikzpicture}
    \end{minipage}\hfill
    \begin{minipage}[t]{0.5\linewidth}
        \vspace{0pt}
        \begin{tikzpicture}[scale=0.1]
            \tikzset{
                bitnode/.style={
                    rotate=80
                },
                simnode/.style={
                    scale=0.1,
                    anchor=south,
                    fill=green,
                    minimum width=4.2cm,
                    draw=black,
                    very thin,
                },
                ibmnode/.style={
                    scale=0.1,
                    anchor=south,
                    fill=blue,
                    minimum width=4.2cm,
                    draw=black,
                    very thin,
                }
            }

            \node (a) at (-2,65) {\textbf{5.}};

            \node[bitnode] (b1) at (0,-1) {010001001};
            \node[bitnode] (b2) at (9,-1) {010001010};
            \node[bitnode] (b3) at (18,-1) {010001100};
            \node[bitnode] (b4) at (27,-1) {010100001};
            \node[bitnode] (b5) at (36,-1) {010100100};
            \node[bitnode] (b6) at (45,-1) {011001100};
            \node[bitnode] (b7) at (54,-1) {rest};

            \draw (-6.5,9) --++(66.5,0);

            \draw (0,8.5) --++(0,1);
            \draw (9,8.5) --++(0,1);
            \draw (18,8.5) --++(0,1);
            \draw (27,8.5) --++(0,1);
            \draw (36,8.5) --++(0,1);
            \draw (45,8.5) --++(0,1);
            \draw (54,8.5) --++(0,1);

            \draw (-6,9) --++(0,72);
            \draw[gray,dashed] (-6,24) --++(65,0);
            \draw (-6.5,24) --++(1,0);
            \draw[gray,dashed] (-6,39) --++(65,0);
            \draw (-6.5,39) --++(1,0);
            \draw[gray,dashed] (-6,54) --++(65,0);
            \draw (-6.5,54) --++(1,0);
            \draw[gray,dashed] (-6,69) --++(65,0);
            \draw (-6.5,69) --++(1,0);

            \node[simnode,minimum height=180mm] (s1) at (-2.1,9) {};
            \node[ibmnode,minimum height=70mm] (i1) at (2.1,9) {};

            \node[ibmnode,minimum height=31mm] (i2) at (11.1,9) {};

            \node[simnode,minimum height=542mm] (s3) at (15.9,9) {};
            \node[ibmnode,minimum height=190mm] (i3) at (20.1,9) {};

            \node[simnode,minimum height=51mm] (s4) at (24.9,9) {};

            \node[simnode,minimum height=140mm] (s5) at (33.9,9) {};
            \node[ibmnode,minimum height=63mm] (i5) at (38.1,9) {};

            \node[ibmnode,minimum height=98mm] (i6) at (47.1,9) {};

            \node[simnode,minimum height=111mm] (s7) at (51.9,9) {};
            \node[ibmnode,minimum height=572mm] (i7) at (56.1,9) {};

            \node[fill=green,minimum width=6mm,draw=black,very thin,anchor=west] (simlegend) at (-2,77.5) {};
            \node[anchor=west] (simlegendtext) at (4,77.5) {simulator};
            \node[fill=blue,minimum width=6mm,draw=black,very thin,anchor=west] (ibmlegend) at (-2,73.5) {};
            \node[anchor=west] (ibmlegendtext) at (4,73.5) {IBM Q System One};
            \draw[draw=black,rounded corners] (-4.5,71) rectangle ++ (47,9);
        \end{tikzpicture}
    \end{minipage}
    \caption{\label{figure:LastSteps}
        Continued numerical experiment for the OSSP(1,3,3) instance.
        The two subplots show the distribution of sampling 1024 times from the quantum computer, resp. the simulated state, after four and five iterations of sampled gradient descent.
        Each plot depicts the counts for the six most sampled bit strings in lexicographical order and collects all remaining counts into a ``rest'' state.
        The overlap with the local minimum continues to increase with the \textbf{4.} VQA-step.
        This trend is stronger in the noisy simulation where more than half of all samples after the final \textbf{5.} step are from the optimal state.
        On the quantum device, the optimal state is at least the most sampled state of all computational basis states, but fails to concentrate more than roughly five percent of all samples in itself.}
\end{figure*}

\textbf{\autoref{figure:FirstSteps}} and \textbf{\autoref{figure:LastSteps}} show the sampling results obtained after each round of the sampled gradient descent.
After five iterations we indeed observe a dominating sampling of the local optimum as well as increased overlap with computational basis states that have a small Hamming distance to the local optimum.
There remains, however, a noisy background both in the simulation as well as on the IBM Q System One.
While the local optimum is clearly
\newpage distinguishable in the simulation, the real quantum device introduces so much noise that the noisy background dominates the slight, but visible improvements in sampling the local optimum.
The mismatch in magnitude of noise indicates a too optimistic classical noise model.
At the same time, however, the data obtained from this simulation hints at what capabilities can be unlocked with improved quantum hardware components.
\newpage

While the performance of the IBM Q System One does not suffice to tackle large-scale OSSP instances---neither in terms of available qubits nor in terms of feasible circuit depths---this proof of principle showcases that all the components of our VQA are already implementable and executable on today's quantum hardware.
Quantitatively extrapolating to larger and less error-prone quantum devices based on a single numerical experiment is, of course, not meaningful.
However, we can discuss the effect of improved qubit counts and gate fidelities versus larger instance sizes qualitatively:
First of all, the number of available qubits is a hard upper limit for the accessible problem sizes.
In general, encoding an OSSP($M, T, J$) instance requires $M T J$ qubits;
the busy case $M T = J$ therefore requires $J^{2}$ qubits.
The quadratically increasing instance size is well matched by the recent efforts to increase the number of qubits such that larger instances with tens of jobs can be encoded.
Second, coherence time and gate fidelities massively influence the performance of any quantum algorithm.
Both factors generally do not respect the problem's feasibility structure so that long noisy circuits tend to exit the feasible subspace if its orthogonal complement is large.
This is typically the case for hard constrained COPs such as the OSSP.
Concretely for the busy case, the number of bit strings and hence the Hilbert space dimension of the encoded problem are given by $2^{J^{2}}$.
In comparison, the feasible set ``only'' comprises $J!$ solutions.
For large $J$, random noise is therefore far more likely to steer the parameterized states outside of the feasible subspace.

\section{\label{section:Conclusion}Conclusion}

In this paper we presented a general approach for characterizing the feasibility structure of open-shop scheduling problems.
We utilized the constraint graph model as a general interplay between graph and group theory for determining symmetries for certain problem instances.
This additional perspective allowed us to find feasibility-preserving mappings as graph automorphisms.
We calculated the entire group of such feasibility-preserving functions and proved that it is isomorphic to the automorphism group of the constraint graph.
This is a very strong statement which unfortunately is not transferable to other types of job-shop scheduling problems.
For example, for a generic \textit{flexible} job-shop instance, where each job is subdivided into an ordered sequence of operations, these ordering constraints introduce additional edges that break certain symmetries, rendering the constraint graph's automorphism group significantly smaller.
The remaining symmetries can still be incorporated into QAOA-mixers but have to be supplemented with additional elements that do not correspond to classical bit permutations.

Let us emphasize one more time that the busy OSSP with $J$ jobs is equivalent to optimizing over the symmetric group $S_{J}$.
We addressed this type of problem by representing the symmetric group with (products of) parameterized exponentials of SWAP gates.
Here, the discussed transitive action of the group on the solution set guarantees that all possible solutions can be reached by suitable parameter adjustments.
Via a decomposition of arbitrary group elements into a sequence of $J (J - 1)^{2} / 2$ simple transitions, we were able to prove that $J (J - 1)^{2} / 2$ variational parameters are sufficient to reach all feasible OSSP solutions, including the optimum.
In principle, we can generalize this procedure to an arbitrary (finite) group $G$, acting transitively on a given set of solutions $S$.
Then, the permutation representation would still yield permutation operators on the qubit space, which, however, will not correspond to actual qubit permutations anymore.
It would be interesting to characterize in the future the obtained operators and the performance of our algorithm for more exotic cases such as the mentioned perfect matching problem, XOR-SAT, and conservation-constrained network flow.
However, operators that do not respect the tensor product structure of $\hil$ will be generally very difficult to implement.
Additionally, problems with inequality constraints seem to have significantly less exploitable feasibility structure so that the group of (unconditionally) feasibility-preserving operations often ends up to be trivial.
For those kind of problems, additional techniques may be employed to complement our approach, which works well for equality constraints.
The ultimate goal would be to create a toolbox for hardcoding arbitrary constraints and to extend those methods also to mixed-integer and continuous optimization problems.
This would allow to also tackle more complicated, but also more realistic logistic problems like transit scheduling~\cite{Owais2016MultiObjectiveTransitRouteNetworkDesignAsSetCoveringProblem,Owais2022FrequencyBasedTransitAssignmentModelsGraphFormulationStudy}.

Moreover, while the cubic upper bound on the number of required variational parameters constitutes a strong theoretical results, its practical implications ultimately hinge on the employed classical optimizer's ability to find high-quality parameter values.
For larger problem instances, simultaneous optimization of cubically many parameters will quickly decline in performance due to exponentially increasing numbers of local minima and flatness of the parameter landscape (barren plateaus).
However, established frameworks for enhancing the parameter optimization such as warm starts, layer-wise optimization schedules, and reformulations as easier problems are also applicable to our method.
Studying their impact on the practicality of our proposed algorithmic blueprint remains an interesting open research question.


\acknowledgments{We thank Michael Cuntz, Tim Heine, Max Hess, Jan Rasmus Holst, Andreea-Iulia Lefterovici, Lauritz van Luijk, Tobias J.\ Osborne, Lilly Palackal, Leonhard Richter, Marco Tomamichel, and Timo Ziegler for helpful discussions.
    The authors especially thank and are deeply indebted to Benjamin Sambale for many long and fruitful discussions.}

\funding{This research was sponsored by the BMBF under the projects ATIQ and QuBRA, by the state of Lower Saxony and the Volkswagen Foundation under the Quantum Valley Lower Saxony, and by the state of Baden-Württemberg under the project SEQUOIA.}

\authorcontributions{Conceptualization, L.B. and G.K.; methodology, L.B. and G.K.; software, L.B. and G.K.; validation, L.B., G.K., C.T., and R.S.; formal analysis, L.B. and G.K.; investigation, L.B. and G.K.; resources, C.T.; data curation, L.B.; writing---original draft preparation, L.B., G.K., C.T., and R.S.; writing---review and editing, L.B.; visualization, L.B.; supervision, C.T. and R.S.; project administration, C.T. and R.S.; funding acquisition, C.T. and R.S.
    All authors have read and agreed to the published version of the manuscript.}

\conflictsofinterest{The authors declare that they have no competing interests.}

\dataavailability{All data supporting the findings of this publication are available within this article.}

\history{}

\PublishersNote

\cright

\end{document}